\documentclass{jpp}

\usepackage{subfig}
\usepackage{graphicx}
\usepackage{float}
\usepackage[labelfont=bf, textfont=it]{caption}
\usepackage{hyperref}
\usepackage{xcolor}
\usepackage{booktabs}
\usepackage{amsmath}

\newcommand{\er}{\mathbf{e}_R}
\newcommand{\ez}{\mathbf{e}_Z}
\newcommand{\ephi}{\mathbf{e}_\varphi}
\newcommand{\dr}{\partial_R}
\newcommand{\dz}{\partial_Z}
\newcommand{\dph}{\partial_\varphi}
\newcommand{\pdr}{\partial_R}
\newcommand{\pdz}{\partial_Z}
\newcommand{\pdp}{\partial_\varphi}

\begin{document}

\title{Two-fluid boundary turbulence simulations in reversed field pinch plasmas}

\author{M. Giacomin\aff{1,2}
  \corresp{\email{maurzio.giacomin@unipd.it}},
  B. Momo\aff{2}, I. Predebon\aff{2,3}, N. Vianello\aff{2,3} and  M.~Zuin\aff{2,3}}

\affiliation{\aff{1} Department of Physics and Astronomy ``G. Galilei'', University of Padua, 35121 Padua, Italy.
\aff{2} Consorzio RFX (CNR, ENEA, INFN, Università di Padova, Acciaierie Venete SpA), 35127, Padua, Italy.
\aff{3} Istituto per la Scienza e la Tecnologia dei Plasmi (ISTP), CNR, Padua, Italy.}

\maketitle

\begin{abstract}
Turbulent transport in magnetic confinement fusion devices governs the overall plasma confinement properties and regulates the plasma–material interaction at the first wall. In the plasma boundary, turbulence is typically investigated through three-dimensional two-fluid flux-driven turbulence simulations. In this work, the GBS boundary turbulence code is extended to enable turbulence simulations in reversed field pinch configurations, encompassing the reversal surface and an arbitrary level of magnetic chaos. The differential operators implemented in the code are modified to avoid the approximations of large-aspect ratio and weak poloidal magnetic field. Three-dimensional Poisson and Ampère solvers are implemented to allow for turbulence simulations in conditions of partially or fully disrupted magnetic flux surfaces. This modified version of the GBS code is then applied to simulate turbulence in the boundary of RFX-mod reversed field pinch plasmas. Turbulent eddies across the reversal surface show properties similar to those typically found in tokamak boundary turbulence simulations.
Despite the good agreement found with experimental measurements, these simulations reveal a significant limitation of the fluid-based turbulence modeling of the edge region in reversed field pinch plasmas, which arises from the intrinsically short parallel connection length. This conclusion is also supported by a linear gyrokinetic analysis that identifies trapped electron modes as the dominant microinstability in this region.
   
\end{abstract}

\section{Introduction}
\label{sec:intro}

The Reversed Field Pinch (RFP) is a self-organized toroidal magnetic configuration where the toroidal magnetic field, modified by the paramagnetic response of the plasma, reverses its direction near the first wall. 
The safety factor $q=rB_\varphi/(RB_\theta)$ ($r$ and $R$ are the minor and major radii, $B_\theta$ and $B_\varphi$ are the poloidal and toroidal magnetic field components, respectively) of RFP plasmas decreases monotonically with radius, remaining below unity at its maximum value. This condition  promotes strong magnetohydrodynamic (MHD) activity with resonant kink-resistive and tearing modes developing over the full radius. These modes originate magnetic islands that, when they overlap, lead to magnetic chaos, which is the primary cause of cross-field heat transport in RFP plasmas~\citep{marrelli2021}.
The degree of magnetic chaos in RFP plasmas depends on the Lundquist number. At low Lundquist number, RFP plasmas are in the so-called multiple helicity (MH) state, in which several resonant modes have similar amplitude, leading to strong magnetic chaos and poor confinement. At high Lundquist number, RFP plasmas show a spontaneous transition to a state where a single resonant MHD mode dominates the spectrum, leading to a reduction of the magnetic chaos intensity and better performance~\citep{lorenzini2009}. In this state, named as quasi-single helicity (QSH), the plasma and the magnetic field self-organize into a helical shape with toroidal periodicity set by the dominant resonant mode. 

The experimental observation of the QSH state has motivated comprehensive investigations of turbulence in RFP plasmas as a potential mechanism governing energy confinement, in direct analogy with the established role of turbulence in tokamak plasmas. These investigations have been conducted following methodological and conceptual frameworks closely aligned with the extensive turbulence research performed in the core of tokamak and stellarator configurations.
Gyrokinetic simulations, carried out in RFP plasmas, have identified non-negligible contributions to turbulent transport arising from ion temperature gradient, electron temperature gradient, trapped electron mode, and microtearing mode instabilities~\citep{predebon2010,carmody2013,predebon2015,williams2017,predebon2019,jitsuk2026}. These findings have been corroborated by experimental observations~\citep{ren2011,zuin2013,thuecks2017,duff2018}.
These simulations have been carried out assuming the existence of well-conserved magnetic flux surfaces and avoiding the plasma region of vanishing toroidal magnetic field. 

Recent global gyrokinetic simulations have demonstrated the importance of retaining tearing-mode-driven magnetic islands and magnetic chaos for an accurate description of turbulent transport in RFP plasmas~\citep{jitsuk2026}. In the edge region of RFP configurations, the interaction between magnetic chaos and turbulence is further complicated by the reversal of the toroidal magnetic field direction. Magnetic chaos and magnetic field reversal pose several numerical challenges. Only very recently, this complex dynamics has been investigated by means of global flux-driven turbulence simulations~\citep{giacomin2026}. 
In that investigation, the GBS code has been applied to model turbulence in a RFX-mod RFP plasma configuration including the reversal surface, with the specific objective of disentangling the role of stochastic transport and turbulence to the self-organized formation of the ambipolar radial electric field.

The numerical analysis presented in this work is performed using a modified version of the GBS code~\citep{Ricci2012,giacomin2022gbs}. 
This work extends the initial numerical investigation of RFP boundary turbulence of \citet{giacomin2026} by: (i) providing a detailed description of the numerical implementation and corresponding verification of the three-dimensional solvers and generalized differential operators; (ii) quantifying the impact of plasma resistivity and $\beta$ on turbulence dynamics and equilibrium profiles in a RFX-mod RFP plasma; and (iii) assessing the applicability of the theoretical framework for tokamak boundary turbulence developed in \citet{giacomin2022turbulent} to RFP configurations. 
The RFP magnetic equilibrium considered here corresponds to the QSH case of \citet{giacomin2026}.
 
This works is organized as follows. Section~\ref{sec:gbs} presents the main changes implemented in the GBS code to enable turbulence simulations across the reversal surface of RFP plasmas. An overview of the  RFP simulations considered in this work is reported in section~\ref{sec:simulations}. Section~\ref{sec:analysis} is dedicated to the analysis of the simulation results, highlighting the main analogies and differences with respect to the tokamak boundary turbulence regimes identified in \citet{giacomin2022turbulent} and discussing the main limitations behind a two-fluid approach in modeling boundary turbulence in RFP plasmas.
The results of a local linear gyrokinetic analysis in the plasma edge of the considered RFP magnetic equilibrium is reported in section~\ref{sec:gyrokinetic}.
Conclusions follow in section~\ref{sec:conclusion}.

\section{The modified GBS code for RFP simulations}
\label{sec:gbs}

GBS is a global, three-dimensional, flux-driven, two-fluid turbulence code that evolves the drift-reduced Braginskii equations~\citep{braginskii1965,Zeiler1997} coupled to a multi-species kinetic model for neutral particle dynamics~\citep{mancini2023}. The simulation domain is discretized on a non-field-aligned cylindrical grid ($R$, $\varphi$, $Z$), where $R$ is the torus major radius, $\varphi$ is the toroidal angle and $Z$ is the vertical coordinate. Thanks to its non-field-aligned grid, GBS can simulate non-trivial magnetic configurations~\citep{giacomin2020snow}, including non-axisymmetric magnetic equilibria~\citep{coelho2022}. On the other hand, the GBS differential operators have been implemented considering the limits of weak poloidal magnetic field, $B_p/B\ll 1$, and large aspect-ratio, $a/R\ll 1$ ($a$ is the torus minor radius). Under these assumptions, the Poisson and Ampère equations are formulated as two-dimensional problems, with the corresponding solvers operating on poloidal planes. 
The approximation $B_p/B\ll 1$ is clearly violated in RFP plasmas. Near the reversal surface, the total magnetic field is almost entirely in the poloidal plane, i.e., $B_p\simeq B$ and $B_p/B_\varphi \gg 1$, making the main version of GBS unsuitable for RFP simulations. 
This section summarizes the main modifications applied to GBS to enable turbulence simulations of RFP plasmas. Further numerical details on GBS can be found in \citet{giacomin2022gbs}.

\subsection{The physical model}

The multi-species neutral model is independent of the magnetic geometry and it does not require any modification. Although included in this code version, the coupling to neutrals is neglected here for simplicity.
The plasma model is similar to the one described in \citet{giacomin2022gbs}, with a few modifications that are described in this section. The plasma model equations implemented in the GBS version used herein are (assuming a single ion species with $Z_i=1$)
\begin{align}
\label{eqn:density}
\frac{\partial n}{\partial t} =& -\frac{1}{B}[\phi,n]+\frac{2}{eB}\Bigl[\mathcal{C}(p_e)-en\mathcal{C}(\phi)\Bigr] 
-\nabla_{\parallel}(n v_{\parallel e}) + D_n\nabla_{\perp}^2 n\nonumber\\
&+s_n+\nu_\text{iz}n_\text{n}-\nu_\text{rec}n\, ,\\
\label{eqn:vorticity}
\frac{\partial \Omega}{\partial t} =& -\frac{1}{B}\nabla \cdot \Bigl(\frac{1}{\Omega_{ci}}[\phi,\boldsymbol{\omega}]\Bigr) - \nabla \cdot \Bigl( \frac{v_{\parallel i}}{\Omega_{ci}}\nabla_\parallel \boldsymbol{\omega}\Bigr)+ \frac{1}{e}\nabla_{\parallel}j_{\parallel}\nonumber\\
&+ \frac{2}{eB} \mathcal{C}(p_e + p_i) + D_{\Omega}\nabla_\perp^2 \Omega-\frac{n_\text{n}}{n}\nu_\text{cx}\Omega\, ,\\
\label{eqn:electron_velocity}
\frac{\partial U_{\parallel e}}{\partial t} =& -\frac{1}{B}[\phi,v_{\parallel e}]  + \frac{e}{m_e}\Bigl( \frac{j_\parallel}{\sigma_\parallel}+\nabla_\parallel\phi -\frac{1}{en}\nabla_\parallel p_e-\frac{0.71}{e}\nabla_\parallel T_e -\frac{4\eta_{0e}}{3en}\nabla_\parallel^2 v_{\parallel e}\Bigr) \nonumber\\
&- v_{\parallel e}\nabla_\parallel v_{\parallel e} + D_{v_{\parallel e}}\nabla_\perp^2 v_{\parallel e}+\frac{n_\text{n}}{n}(\nu_\text{en}+2\nu_\text{iz})(v_{\parallel n}-v_{\parallel e})\,,\\
\label{eqn:ion_velocity}
\frac{\partial v_{\parallel i}}{\partial t} =& -\frac{1}{B}[\phi,v_{\parallel i}] - v_{\parallel i}\nabla_\parallel v_{\parallel i} - \frac{1}{m_i n}\nabla_\parallel(p_e+ p_i) -\frac{4\eta_{0i}}{3m_in}\nabla_\parallel^2 v_{\parallel i} + D_{v_{\parallel i}}\nabla_\perp^2 v_{\parallel i}\nonumber\\
& +\frac{n_\text{n}}{n}(\nu_\text{iz}+\nu_\text{cx})(v_{\parallel n}-v_{\parallel i})\,,\\
\label{eqn:electron_temperature}
\frac{\partial T_e}{\partial t} =& -\frac{1}{B}[\phi,T_e] - v_{\parallel e}\nabla_\parallel T_e 
+ \frac{2}{3}T_e\Bigl[0.71\frac{\nabla_\parallel j_\parallel}{en} - \nabla_\parallel v_{\parallel e}\Bigr] + \nabla_\parallel (\chi_{\parallel e}\nabla_\parallel T_e) + D_{T_e}\nabla_\perp^2 T_e \nonumber\\
&+ \frac{4}{3}\frac{T_e}{eB}\Bigl[\frac{7}{2}\mathcal{C}(T_e)+\frac{T_e}{n}\mathcal{C}(n)-e\mathcal{C}(\phi)\Bigr]  + s_{T_e}-\frac{n_\text{n}}{n}\nu_\text{en}m_e\frac{2}{3}v_{\parallel e}(v_{\parallel n}-v_{\parallel e})\nonumber\\
&- \frac{4}{3}\frac{m_e}{m_i}\frac{1}{\tau_e} (T_e- T_i) + \frac{n_\text{n}}{n}\nu_\text{iz}\biggl[-\frac{2}{3}E_\text{iz}-T_e+m_e v_{\parallel e}\Bigl(v_{\parallel e}-\frac{4}{3}v_{\parallel n}\Bigr)\biggr]  \,,
\end{align}
\begin{align}
\label{eqn:ion_temperature}
\frac{\partial T_i}{\partial t} =& -\frac{1}{B}[\phi,T_i] - v_{\parallel i}\nabla_\parallel T_i + \frac{4}{3}\frac{T_i}{eB}\Bigl[\mathcal{C}(T_e +\frac{T_e}{n}\mathcal{C}(n)-e\mathcal{C}(\phi)\Bigr] - \frac{10}{3}\frac{T_i}{eB}\mathcal{C}(T_i) \nonumber\\
&+ \frac{2}{3}T_i\Bigl[(v_{\parallel i}-v_{\parallel e})\frac{\nabla_\parallel n}{n} -\nabla_\parallel v_{\parallel e}\Bigr] 
 + \nabla_\parallel (\chi_{\parallel i}\nabla_\parallel T_i) + D_{T_i}\nabla_\perp^2 T_i + s_{T_i} \nonumber\\
 &+ \frac{4}{3} \frac{m_e}{m_i}\frac{1}{\tau_e} (T_e- T_i) + \frac{n_\text{n}}{n}(\nu_\text{iz}+\nu_\text{cx})\Bigl[T_n-T_i+\frac{1}{3}(v_{\parallel n}-v_{\parallel i})^2\Bigr]\,,
\end{align}
which are coupled to the Poisson and Ampère equations,
\begin{align}
\label{eqn:poisson}
\nabla \cdot \Bigl( \frac{n}{B\Omega_{ci}} \nabla_\perp \phi\Bigr) =&\ \Omega- \nabla\cdot \Bigl(\frac{1}{m_i\Omega_{ci}^2}\nabla_\perp p_i\Bigr)\,,\\
\label{eqn:ampere}
\biggl( \nabla_\perp^2 - \frac{e^2\mu_0}{m_e}n\biggr)v_{\parallel e}=&\ \nabla_\perp^2 U_{\parallel e} - \frac{e^2\mu_0}{m_e}n v_{\parallel i}\,.
\end{align}
In Eqs.~\eqref{eqn:density}--\eqref{eqn:ampere}, $n$ is the plasma density ($n=n_e=n_i$ for a single ion species with $Z_i=1$), $T_e$ and $T_i$ are the electron and ion temperatures, $p_e=nT_e$ and $p_i=nT_i$ are the electron and ion pressures, $U_{\parallel e} = v_{\parallel e} +  e \psi/m_e$ is the sum of the electron inertia and the electromagnetic induction contributions, $v_{\parallel i}$ is the ion parallel velocity, $\Omega = \nabla\cdot(\boldsymbol{\omega}/\Omega_{ci}) = \nabla \cdot \bigl[(n \nabla_\perp\phi + \nabla_\perp p_i/e)/(\Omega_{ci}B)\bigr]$ is the scalar vorticity, $\Omega_{ci}=eB/m_i$ is the ion cyclotron frequency, $\nu_\textrm{iz}$, $\nu_\textrm{rec}$, $\nu_\textrm{cx}$ and $\nu_\textrm{en}$ are the ionization, recombination, charge-exchange and electron-neutral collisions frequencies, respectively (see \citet{giacomin2022gbs} for details on the plasma-neutral coupling), $E_\textrm{iz}$ is the ionization energy, $\sigma_\parallel = 1.96 n e^2 \tau_e/m_e$ is the parallel electric conductivity, $\chi_e$ and $\chi_i$ are the electron and ion parallel thermal conductivities, as defined in the Braginskii's model~\citep{braginskii1965}.
 
The spatial operators in Eqs.~\eqref{eqn:density}--\eqref{eqn:ampere} are the $\mathbf{E}\times\mathbf{B}$ convective term (Poisson brackets),
\begin{equation}
\label{eqn:pb}
    [\phi,f]=\mathbf{b}\ \cdot\ \bigl(\nabla \phi \times \nabla f\bigr)\,,\\
\end{equation}
the curvature operator,
\begin{equation}
\label{eqn:curv}
    \mathcal{C}(f)=\frac{B}{2}\Bigl(\nabla \times \frac{\mathbf{b}}{B}\Bigr)\cdot \nabla f\,,\\
\end{equation}
the parallel gradient, which includes the electromagnetic flutter contribution, 
\begin{equation}
\label{eqn:gradpar}
    \nabla_\parallel f=\mathbf{b}\cdot\nabla f + \frac{1}{B}[\psi,f]\,,\\
\end{equation}
and the perpendicular Laplacian,
\begin{equation}
\label{eqn:lapl}
    \nabla_\perp^2 f=\nabla\cdot\bigl[(\mathbf{b}\times\nabla f)\times\mathbf{b}\bigl]\,,\\
\end{equation}
where $f$ is a general scalar function. 

Eqs.~\eqref{eqn:density}--\eqref{eqn:ampere} are implemented in GBS in their dimensionless form, where $n$ is normalized to the reference density $n_0$, $T_e$ and $T_i$ are normalized to the reference electron temperature $T_{e0}$, $v_{\parallel e}$ and $v_{\parallel i}$ are normalized to the reference sound speed $c_{s0}=\sqrt{T_{e0}/m_i}$, the magnetic field is normalized to the reference magnetic field strength $B_0$, the electrostatic potential is normalized to $T_{e0}/e$, $\psi$ is normalized to $\rho_{s0}B_0$, with $\rho_{s0}=c_{s0}/\Omega_{ci0}$, perpendicular and parallel lengths are normalized to $\rho_{s0}$ and $R_0$, respectively, with $R_0$ the machine major radius, and the time is normalized to $R_0/c_{s0}$. 
The reference electron plasma $\beta$, $\beta_{e0}=2\mu_0n_0T_{e0}/B_0^2$, and the normalized Spitzer resistivity, $\nu=e^2n_0R_0/(m_ic_{s0}\sigma_\parallel) = \nu_0T_e^{-3/2}$, stem from the normalization process, with
\begin{equation}
\label{eqn:nu0}
    \nu_0 = \frac{4\sqrt{2\pi}}{5.88}\frac{e^4}{(4\pi \epsilon_0)^2}\frac{\sqrt{m_e}R_0n_0\lambda}{m_ic_{s0}T_{e0}^{3/2}}
\end{equation}
the reference resistivity, which depends on the reference density and reference electron temperature ($\lambda$ is the Coulomb logarithm).

The main difference between the physical model considered in the present work and the one described in \citet{giacomin2022gbs} is related to the spatial dependence of the modulus of the magnetic field, $B$. In fact, under the assumption of weak poloidal magnetic, the magnetic field modulus is approximated to $B^2 = B_\varphi^2 + B_p^2 = B_\varphi^2(1+B_p^2/B_\varphi^2)\simeq B_\varphi^2$, namely $B\simeq |B_\varphi|$. In the large aspect-ratio limit and by considering $B_0$ as the magnetic field strength at the magnetic axis, the modulus of the magnetic field is approximated by $|B_\varphi| = B_0R_0/R\simeq B_0$. This leads to $B\simeq B_0$ and $\Omega_{ci} \simeq \Omega_{ci0}$. 
Consequently, the magnetic field dependence in the scalar vorticity is neglected in \citet{giacomin2022gbs}, i.e. $\Omega \propto \nabla \cdot (n \nabla_\perp\phi + \nabla_\perp p_i/e)$. This approximation simplifies the implementation of the vorticity equation (see Eq.~\eqref{eqn:vorticity}) and Poisson law (see Eq.~\eqref{eqn:poisson}). In addition, the factor $1/B$ in front of the Poisson brackets in Eqs.~\eqref{eqn:density}--\eqref{eqn:ion_temperature} is neglected in \citet{giacomin2022gbs} when these equations are written in dimensionless form ($B/B_0=1$). These approximations are justified in conventional tokamaks, where $B$ varies weakly within the plasma volume. On the other hand, they do not hold in general, and the spatial dependence of $B$ is therefore retained in the present work.
 
\subsection{The extended GBS operators}

The GBS domain is discretized on a cylindrical uniformly-spaced $(R, \varphi, Z)$ grid. The spatial derivatives are implemented with a fourth-order centred finite differences scheme and the time evolution is provided by a fourth-order Runge-Kutta algorithm. An iterative solver based on the Data Management for Structured Grids (DMDA) of the \texttt{PETSc} library~\citep{petsc-web-page} is used for the implementation of Poisson and Ampère equations.
While the spatial discretization and temporal advancement schemes are identical to those employed in the main GBS version, the differential operators in Eqs.~\eqref{eqn:pb}--\eqref{eqn:lapl} are here derived in their most general form, explicitly avoiding the asymptotic expansions in $\epsilon\equiv r/R_0 \ll 1$ and $\delta \equiv B_p/B \ll 1$.

The magnetic field is written in cylindrical coordinates, $\mathbf{B} = B_R \er + B_Z \ez + B_\varphi \ephi$, where $\er$, $\ez$ and $\ephi$ denote the radial, vertical and toroidal directions, respectively. Whereas the differential operators in \citet{giacomin2022gbs} are expressed in terms of the poloidal flux function $\Psi$ and its spatial derivatives, in the present work they are instead formulated directly in terms of the magnetic field components $B_R$, $B_Z$, $B_\varphi$ and their corresponding spatial derivatives. This choice enables turbulence simulations also in magnetic configurations without well-converted magnetic flux surfaces, such as those with magnetic chaos.

In cylindrical coordinates, the Poisson brackets operator, Eq.~\eqref{eqn:pb}, becomes
\begin{equation}
\begin{split}
    [\phi, f] &= \frac{1}{R}\frac{B_R}{B}(\dph\phi\dz f - \dz\phi\dph f)\\
    & +\frac{1}{R}\frac{B_Z}{B}(\dr\phi\dph f- \dph\phi\dr f)\\
    & +\frac{B_\varphi}{B}(\dz\phi\dr f- \dr\phi\dz f)\,,
\end{split}
\end{equation}
which is normalized to $1/\rho_{s0}^2$. Only the last term is retained in \citet{giacomin2022gbs} when the operator is expanded in $\epsilon$ and $\delta$. As in \citet{giacomin2022gbs}, the Poisson brackets operator is discretized using a fourth-order Arakawa scheme~\citep{peterson2013}.  

\vspace{0.8cm}
The curvature operator, Eq.~\eqref{eqn:curv}, is written as
\begin{equation}
\begin{split}
    \mathcal{C}(f) = \frac{1}{2 B^3 R}\Bigl(&\pdp f  \bigl(\pdz B_R  (B^2-2 B_R ^2)-\pdr B_Z  (B^2-2 B_Z ^2) \\
    &-2 B_R  \bigl(B_Z  (\pdz B_Z -\pdr B_R )+B_\varphi  \pdz B_\varphi \bigr)+2 B_Z B_\varphi  \pdr B_\varphi \bigr) \\
    &+\pdz f  \bigl(-\pdp B_R  (B^2-2 B_R ^2)+B_\varphi (B^2+2 B_R  (\pdp B_\varphi -R\pdr B_R )\\
    &-2 R B_Z  \pdr B_Z) + RB^2 \pdr B_\varphi +2 B_R  B_Z  \pdp B_Z -2 R B_\varphi ^2 \pdr B_\varphi \bigr) \\
    &+\pdr f  \bigl(\pdp B_Z (B^2-2 B_Z ^2)-R\pdz B_\varphi (B^2-2 B_\varphi ^2) \\
    &+B_R  (2 R B_\varphi \pdz B_R  -2 B_Z  \pdp B_R )-2 B_Z  B_\varphi  (\pdp B_\varphi -R\pdz B_Z )\bigr)\Bigr)\,,
\end{split}
\end{equation}
and it is normalized to $1/(R_0\rho_{s0})$. If only the leading order terms in $\epsilon$ and $\delta$ are retained and local current are neglected, as in \citet{giacomin2022gbs}, the normalized curvature operator is drastically simplified to $\mathcal{C}(f) = (B_\varphi/B_0)\partial_Z f$, which corresponds to the curvature associated with the toroidal magnetic field.  

The parallel gradient operator, Eq.~\eqref{eqn:gradpar}, becomes
\begin{equation}
\label{eqn:gradpar_cyl}
    \nabla_\parallel f = \frac{B_R}{B}\pdr f + \frac{B_Z}{B}\pdz f + \frac{B_\varphi}{BR}\pdp f\,,
\end{equation}
and it is normalized to $1/R_0$. The parallel Laplacian operator is obtained by applying twice the $\nabla_\parallel$ operator, leading to 
\begin{equation}
\label{eqn:lapl_par}
\begin{split}
    \nabla_\parallel^2 f = \,& a_R \pdr f + a_Z \pdz f + a_\varphi \pdp f + a_{RR} \pdr^2f +a_{ZZ}\pdz^2f+a_{\varphi\varphi}\pdp^2 f \\&+ a_{RZ}\partial_{RZ}f + a_{R\varphi}\partial_{R\varphi}f+ a_{Z\varphi}\partial_{Z\varphi}f\,,
\end{split}
\end{equation}
where the coefficients $a_R$, $a_Z$, $a_\varphi$, $a_{RR}$, $a_{ZZ}$, $a_{\varphi\varphi}$, $a_{RZ}$, $a_{R\varphi}$, $a_{Z\varphi}$ are given in appendix~\ref{app:coeff}.

The perpendicular Laplacian, Eq.~\eqref{eqn:lapl}, is written as
\begin{equation}
\label{eqn:lapl_perp}
\begin{split}
\nabla_\perp^2 f=\,& c_R \pdr f + c_Z \pdz f + c_\varphi \pdp f + c_{RR} \pdr^2f+c_{ZZ}\pdz^2f+c_{\varphi\varphi}\pdp^2 f \\&+ c_{RZ}\partial_{RZ}f + c_{R\varphi}\partial_{R\varphi}f+ c_{Z\varphi}\partial_{Z\varphi}f\,,
\end{split}
\end{equation}
where the coefficients $c_R$, $c_Z$, $c_\varphi$, $c_{RR}$, $c_{ZZ}$, $c_{\varphi\varphi}$, $c_{RZ}$, $c_{R\varphi}$, $c_{Z\varphi}$ are given in appendix~\ref{app:coeff}. The perpendicular Laplacian is normalized to $1/\rho_{s0}^2$. 
The Poisson equation, Eq.~\eqref{eqn:poisson}, involves a modified perpendicular Laplacian operator of the form $\nabla \cdot (f \nabla_\perp\phi)$, with $f=n/B^2$ in the left-hand side and $f=1/B^2$ in the right-hand side. This operator is written as in Eq.~\eqref{eqn:lapl_perp} with the coefficients given in appendix~\ref{app:coeff}.

At leading order in $\epsilon$ and $\delta$, the perpendicular Laplacian in Eq.~\eqref{eqn:lapl_perp} undergoes a substantial simplification, such that $c_{RR}$ and $c_{ZZ}$ are the only coefficients that remain non-vanishing. Consequently, the Laplacian operator appearing in Poisson and Ampère equations is reduced to a two-dimensional operator acting on the $(R, Z)$ plane. 
However, in the most general case (including RFP plasmas), the coefficients in Eq.~\eqref{eqn:lapl_perp} that multiply the derivatives in the toroidal direction do not vanish, thus requiring the use of three-dimensional  solvers in Eqs.~\eqref{eqn:poisson}--\eqref{eqn:ampere}. This is the major code modification with respect to \citet{giacomin2022gbs}.
The three-dimensional iterative solvers are implemented using the DMDA of the \texttt{PETSc} library. As in the two-dimensional case, the best performance is achieved with the algebraic multigrid preconditioner \texttt{boomerAMG} provided by the \texttt{HYPRE} library~\citep{hypre} and the flexible generalized minimal residual (flexible \texttt{GMRES}) solver~\citep{saad1993}. The preconditioner parameters are adapted to the three-dimensional solver, changing the coarsen type from \texttt{Falgout} to \texttt{HMIS} and increasing the strong threshold from 0.25 to 0.7.

\subsection{Verification and scalability}

The modifications applied to the differential operators and, more critically, the implementation of  three-dimensional solvers necessitate a renewed verification of the code. This verification is performed using the Method of Manufactured Solutions (MMS)~\citep{riva2014}, as described below. 

The MMS is a numerical technique that consists in verifying the correct implementation of a model $M$. The verification is successful if the numerical solution $s_h$ of the discretized model $M_h$ converges to the analytical solution $s$ of the model $M$ as the discretization error $h$ tends to zero, i.e. $e_h = ||s-s_h|| \to h^p$ as $h \to 0$ with $p$ the order of discretization scheme ($p=4$ in GBS since both space and time are discretized with a fourth-order method). 
Since the analytical solution of the model equations is in general unknown, the fundamental idea underlying the MMS is to prescribe an arbitrary analytical function, denoted by \(u\), and to evaluate the corresponding source term \(S = M(u)\), where \(M\) represents the continuous model operator. The manufactured function \(u\) is then, by construction, an exact solution of the modified model \(N\), defined as $N(u) = M(u) - S = 0$.
Since \(S\) can be analytically evaluated, the discrete operators \(N_h\) and $M_h$ are affected by the same discretization error.
Consequently, the verification of the correctness of the discrete formulation \(N_h\) is equivalent to the verification of the discrete model operator \(M_h\).

The verification is carried out by choosing the same numerical setup as in \citet{giacomin2022gbs}, i.e.,  $\rho_*^{-1}= R_0/\rho_{s0} = 100$, $\nu_0=1.0$, $m_e/m_i=1.0$, $\beta_{e0} = 10^{-4}$, and a domain size in the radial and vertical direction of $L_R=37.5\,\rho_{s0}$ and $L_Z=50\,\rho_{s0}$. The magnetic field is defined as
\begin{align}
\label{eqn:mag_ver_1}
    B_R &= \sin\biggl(\frac{2\pi R}{R_\mathrm{max}-R_\mathrm{min}}\biggr)\sin\biggl(\frac{2\pi Z}{Z_\mathrm{max}-Z_\mathrm{min}}\biggr)\cos(\varphi)\\
    B_Z &= \cos\biggl(\frac{2\pi R}{R_\mathrm{max}-R_\mathrm{min}}\biggr)\cos\biggl(\frac{2\pi Z}{Z_\mathrm{max}-Z_\mathrm{min}}\biggr)\cos(\varphi)\\
\label{eqn:mag_ver_2}
    B_\varphi &= 2 + \cos\biggl(\frac{2\pi R}{R_\mathrm{max}-R_\mathrm{min}}\Bigr)\cos\biggl(\frac{2\pi Z}{Z_\mathrm{max}-Z_\mathrm{min}}\biggr)\sin(\varphi)
\end{align}
in such a way that all coefficients of the differential operators are nonzero.
We remark that the verification is independent of the choice of the equilibrium magnetic field, as long as it leads to nontrivial differential operators 
Similarly to \citet{giacomin2022gbs}, the prescribed solutions $u=n, T_e, T_i, \phi, \Omega, v_{\parallel e}, v_{\parallel i}, \psi$ are chosen as
\begin{equation}
    u (R, \varphi, Z) = A_u\bigl[B_u + \sin(C_uZ+\alpha_u)\sin(D_u\varphi+\beta_u)\sin(E_u t + F_u R + \gamma_u) \bigr]
\end{equation}
where $A_u$, $B_u$, $C_u$, $D_u$, $E_u$, $F_u$, $\alpha_u$, $\beta_u$ and $\gamma_u$ are arbitrary constants. The source term is computed by substituting the prescribed solution into Eqs.~\eqref{eqn:density}--\eqref{eqn:poisson}, with the magnetic field given by Eqs.~\eqref{eqn:mag_ver_1}--\eqref{eqn:mag_ver_2}.

\begin{figure}
    \centering
    \subfloat[]{\includegraphics[width=0.47\textwidth]{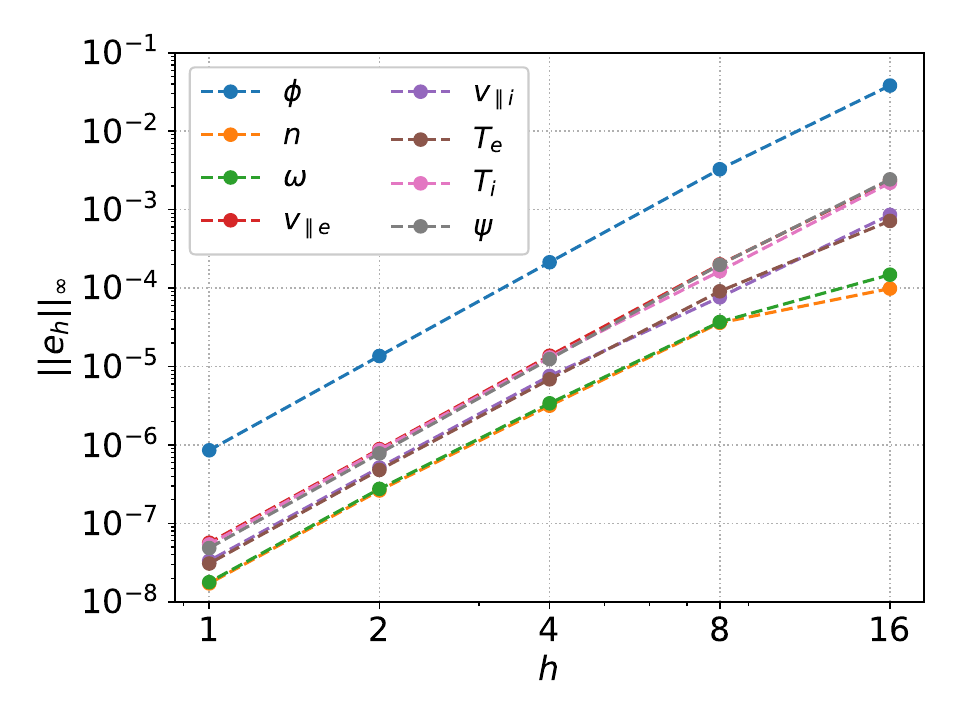}}
    \subfloat[]{\includegraphics[width=0.47\textwidth]{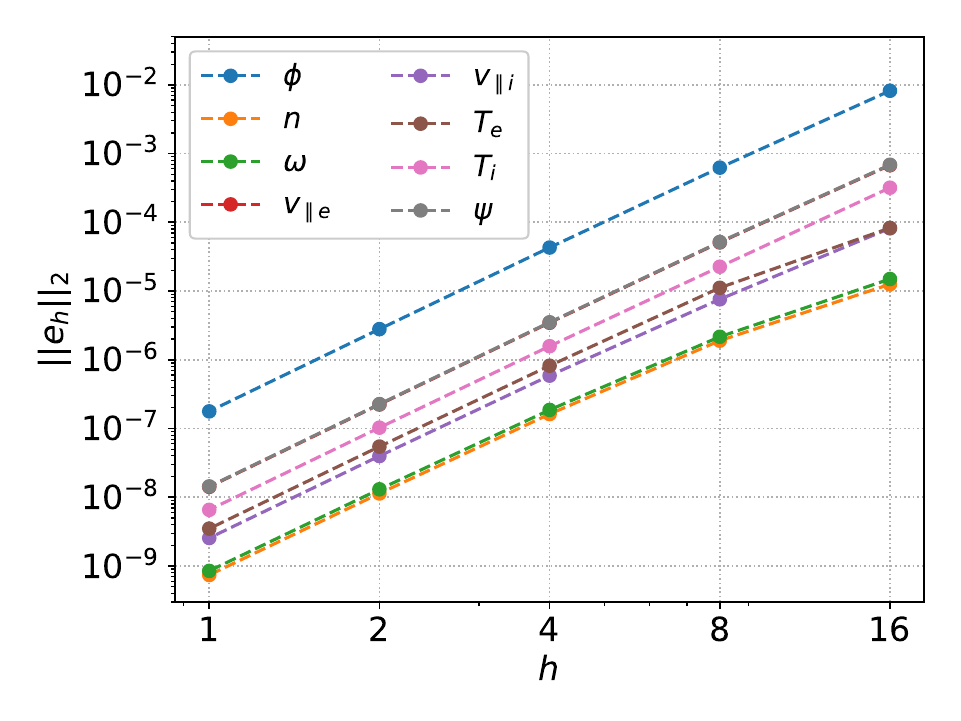}}\\
    \subfloat[]{\includegraphics[width=0.47\textwidth]{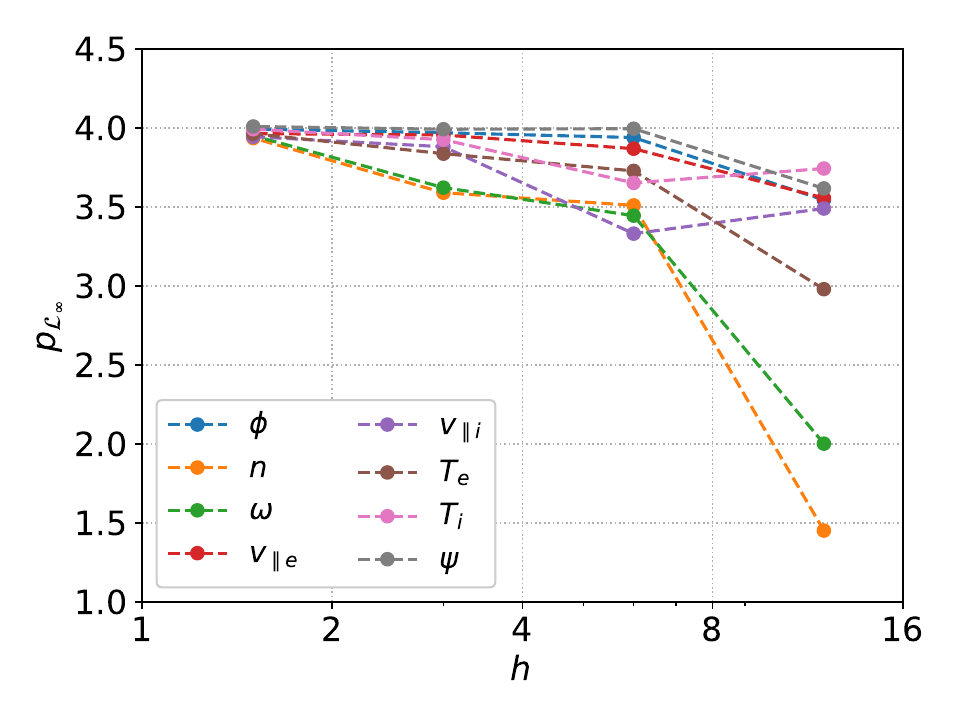}}
    \subfloat[]{\includegraphics[width=0.47\textwidth]{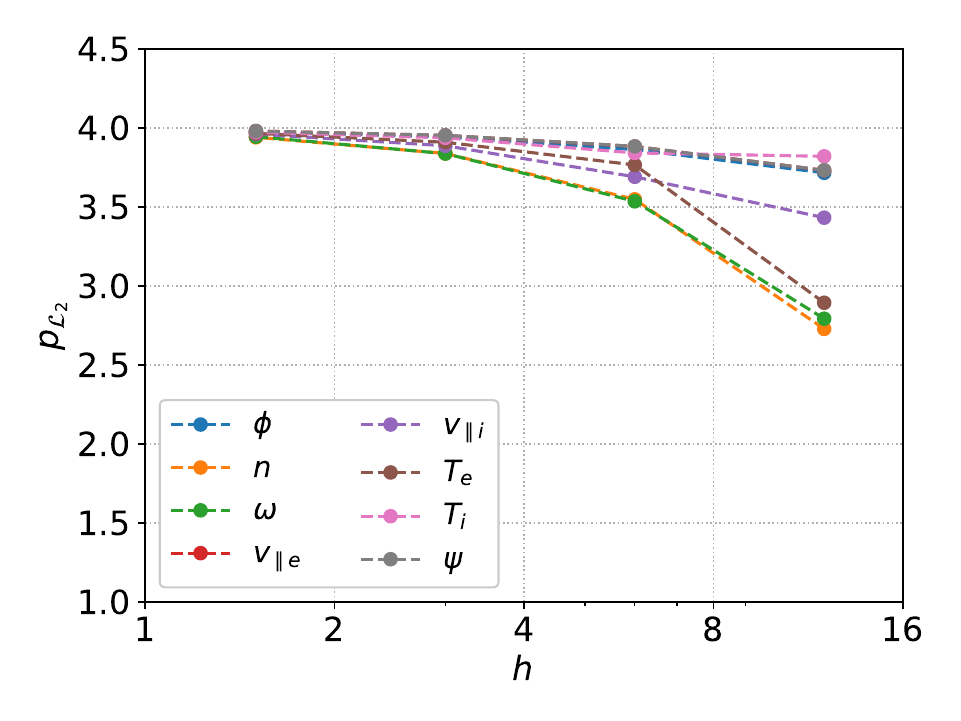}}
    \caption{Results of the verification test. $\mathcal{L}_\infty$ (a) and $\mathcal{L}_2$ (b) norms of the discretization error as functions of $h$.  Order of convergence $p$ for the $\mathcal{L}_\infty$ (c) and the $\mathcal{L}_2$ (d) norms of the discretization errors. The order of convergence is evaluated from two consecutive grid resolutions.}
    \label{fig:verification}
\end{figure}

The verification is performed by considering five grids of increasing spacing, with the finest grid $(N_R^*=128, N_\varphi^*=128, N_Z^* = 128)$ and $dt^*=1.25\times 10^{-5}\,R_0/c_{s0}$. The discretization parameter $h\in \{1, 2, 4, 8\}$ is defined as $h=N_R^*/N_R=N_Z^*/N_Z=N_\varphi^*/N_\varphi= dt/dt^*$. Figure~\ref{fig:verification} shows the $\mathcal{L}_2$ and the $\mathcal{L}_\infty$ norms of the discretization error as a function of $h$, along with the order of convergence $p=\log(e_{rh}/e_h)/\log{r}$, where $r=2$ is the grid refinement. The discretization error scales as $h^4$, hence verifying the implementation of the modified differential operators and three-dimensional solvers.

\begin{figure}
    \centering
    \subfloat[]{\includegraphics[width=0.48\textwidth]{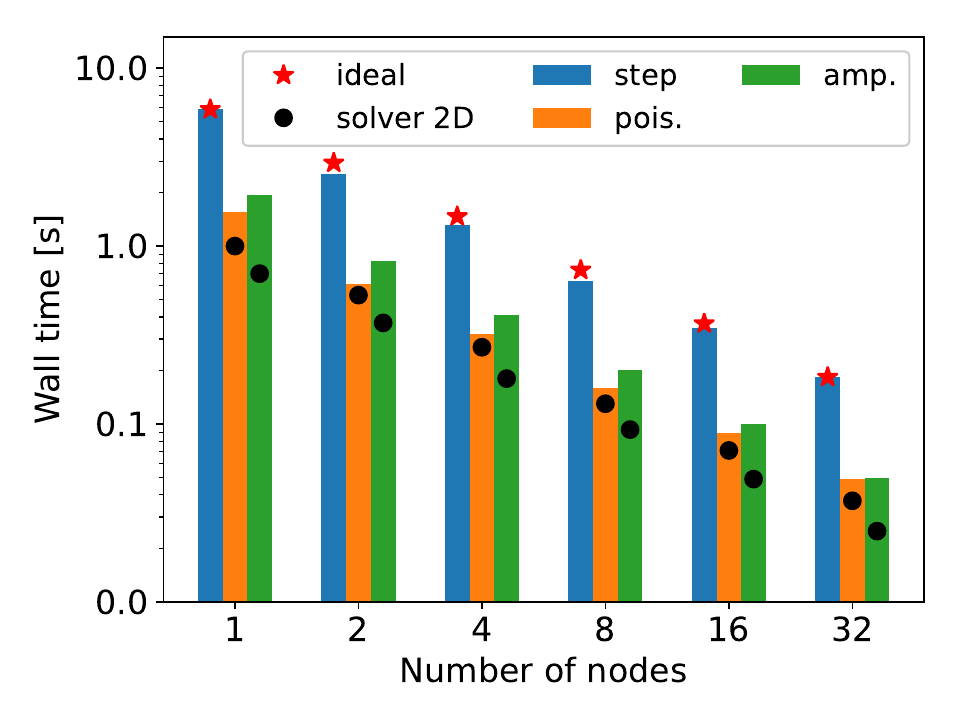}}
    \subfloat[]{\includegraphics[width=0.48\textwidth]{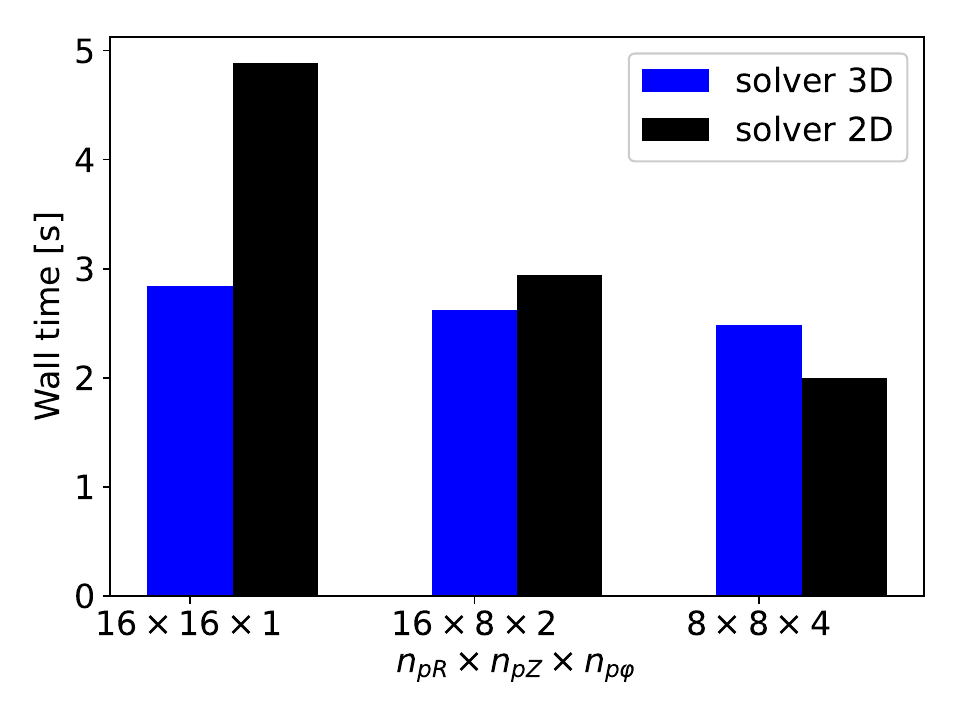}}
    \caption{(a) Strong scalability test performed on Discoverer HPC (2 $\times$ AMD EPYC 7H12 64-Core Processor) considering a grid size of $N_R\times N_\varphi \times N_Z = 160 \times 128 \times 160$ and increasing the number of computing nodes from 1 to 32 (1 node = 128 cores). The wall-clock execution times of the Poisson (orange) and Amperè (green) solvers are shown along with the total wall-clock time required for a single time step (blue). The ideal scaling is denoted by the red markers, while the two-dimensional solver times are indicated by the black markers. (b) Time-to-solution for a single time step for different MPI domain decompositions in GBS simulations employing three-dimensional and two-dimensional solvers. The numbers of MPI tasks distributed along the $R$, $Z$, and $\varphi$ directions are denoted by $n_{pR}$, $n_{pZ}$, and $n_{p\varphi}$, respectively.}
    \label{fig:scaling}
\end{figure}

The parallelization in GBS is implemented through the domain decomposition by using the Message Passing Interface (MPI) library, which, thanks to the use of a uniform cylindrical grid, offers a very efficient scalability up to a large number of MPI tasks. While the code parallelization is not affected by the modifications on the differential operators, the implementation of new iterative solvers may impact the code performance. It is therefore important to test how the wall time of the solvers scales with the number of cores (strong scaling).
The scalability test is carried on Discoverer HPC (2 $\times$ AMD EPYC 7H12 64-Core Processor) considering a grid size of $N_R\times N_\varphi \times N_Z = 160 \times 128 \times 160$, which is a typical grid for a half-size RFX-mod RFP simulation (section~\ref{sec:simulations}).
Figure~\ref{fig:scaling} shows the wall time of one step, computed as time average over 100 steps, as a function of the number of nodes. The time taken by the three-dimensional Poisson and Ampère solvers is also indicated in figure~\ref{fig:scaling}. With the chosen grid, GBS scales almost ideally up to 32 nodes (4092 cores). Importantly, the newly implemented three-dimensional solvers scale well with the number of MPI tasks. For comparison, the time-to-solution related to Poisson and Ampère two-dimensional solvers is also shown in figure~\ref{fig:scaling}(a).
The simulation with two-dimensional solvers is approximately a factor 1.5 faster than the same simulation with three-dimensional solvers, although this factor is sensitive to the computing architecture and MPI domain decomposition, and it can rise above 2.

Because of the intense parallel communication required by  Poisson and Ampère solvers, the GBS version with two-dimensional solvers scales more efficiently in the toroidal direction. Indeed, the parallelization strategy usually adopted is to distribute the MPI tasks within a computing node (or a fraction of it) on the poloidal plane, and increase the number of MPI tasks along $\varphi$ consequently. The implementation of three-dimensional solvers makes the choice of the MPI decomposition irrelevant, as the three directions are equivalent. Indeed, figure~\ref{fig:scaling}(b) shows that the time to solution weakly changes with the chosen MPI decomposition. A larger variation is instead observed if two-dimensional solvers are employed, with the most efficient case corresponding to the one with the largest number of MPI tasks along $\varphi$.

Although the newly implemented three-dimensional iterative solvers do not degrade the excellent scalability properties of GBS, their heavier computational cost makes GBS simulations more expensive. Therefore, it is advisable to use this modified version of GBS only when the approximations $\delta \ll 1$ and/or $\epsilon \ll 1$ are violated or when it is not possible to define a flux function $\Psi$.    

\section{Simulation overview}
\label{sec:simulations}

The modified GBS code is applied here to characterize turbulence and transport in the boundary region of a RFP plasma over a wide range of plasma resistivity and $\beta$.

\subsection{Simulation setup}

The magnetic field considered in this work is reconstructed with the NCT code~\citep{zanca2004} from the RFP plasma discharge \#29187 at $t=209$~ms of the RFX-mod device (minor radius $a=0.46$~m and major radius $R_0=2.0$~m).  This is the same magnetic field considered in \citet{giacomin2026} and it corresponds to a QSH state with plasma current $I_p\simeq 1.5$~MA and shallow reversal. The magnetic field reconstruction includes the axisymmetric equilibrium and all the MHD modes with poloidal mode numbers $m\in\{-1, 0, 1, 2\}$ and toroidal mode numbers up to $n=24$. The dominant mode is $(m, n) = (1, 7)$, which yields an approximately helical magnetic field configuration characterized by a toroidal periodicity of $n = 7$ (see \citet{giacomin2026} for additional details on the magnetic field). 
The Poincaré map of the magnetic field on the $R\varphi$-midplane, depicted in figure~\ref{fig:snapshot}, reveals pronounced magnetic stochasticity, which originates from the overlap of magnetic islands associated with adjacent resonant surfaces. The magnetic island near the wall ($r/a\simeq 1$) is caused by the $m=0,\, n=7$ unstable tearing mode. 

The number of grid points in the radial, toroidal and vertical direction is $N_R=160$, $N_\varphi = 128$ and $N_Z=160$, respectively. The size of the simulation domain is $L_R= 300\,\rho_{s0}$ in the radial direction, $L_Z=300\,\rho_{s0}$ in the vertical direction and $L_\varphi = 2\pi/7$ in the toroidal direction, with $\rho_*^{-1}= R_{s0}/\rho_{s0}=750$. In particular, the quasi-periodicity associated with the dominant MHD mode with toroidal mode number $n=7$ is exploited in this work to restrict the computational toroidal domain to $0 \leq \varphi < 2\pi/7$. 
By considering typical values of electron temperature and magnetic field in the boundary of RFP discharges, $T_{e0} \simeq 60$~eV and $B_0 \simeq 0.6$~T, the reference Larmor radius is $\rho_{s0} \simeq 1.5$~mm. This yields to a poloidal cross-section of size $L_R=L_Z\simeq 0.45$~m. Consequently, the size of the poloidal plane in this set of GBS simulations corresponds to half of the physical size of the RFX-mod device. The major radius in the simulations is also half of the RFX-mod one. Adopting a reduced system size enables comprehensive parameter-space exploration that would otherwise be computationally prohibitive at full scale.

The set of GBS simulations considered here includes a scan on the normalized resistivity, $\nu_0\in \{0.05,\, 0.5,\, 5.0,\, 50\}$ at $\beta_{e0}=10^{-3}$, and a scan on the reference electron $\beta$, $\beta_{e0}\in \{0.001,\, 0.005,\, 0.01,\, 0.05\}$ at $\nu_0=0.05$. 
We note that varying $\nu_0$ is effectively equivalent to changing the collisionality through the reference density $n_0$ (see Eq.~\eqref{eqn:nu0}). The reference quantities are evaluated at the plasma boundary and the experimental values of $\nu_0$ and $\beta_{e0}$ in the considered RFX-mod plasma discharge are approximately $\nu_0 = 0.05$ and $\beta_{e0} = 0.001$.
We highlight that values of $\nu_0$ greater than 1 are unrealistic for a RFX-mod RFP plasma, even in a poorly confined single helicity regime. Similarly, $\beta_{e0}$ values exceeding 0.005 are also not realistic. In this work, GBS simulations employing such non-physical values of $\nu_0$ and $\beta_{e0}$ are carried out solely to test the theoretical framework presented in section~\ref{sec:analysis}.

These parameter scans are performed at fixed temperature and density sources, which are uniform functions along $\varphi$ defined as
\begin{align}
\label{eqn:den_source}
    s_n &= s_{n0} \exp\Bigl[-\frac{(r-r_{n0})^2}{\Delta r_n^2}\Bigr]\,,\\
\label{eqn:temp_source}
    s_{T_e} &= \frac{s_{T0}}{2}\Bigl[1-\tanh\Bigl(\frac{r-r_{T0}}{\Delta r_T}\Bigr)\Bigr]\,,
\end{align}
with $s_{n0} = 0.01\,n_0c_{s0}/R_0$, $r_{n0}=0.5\,a$, $\Delta r_n = 0.1\,a$, $s_{T0}=0.05\,T_{e0}c_{s0}/R_0$, $r_{T0}=0.2\,a$ and $\Delta r_T=0.05\,a$. These parameter values are chosen such that the total particle fueling rate is approximately $10^{21}\,\textrm{particles}/s$ and the applied heating power is $1\,\mathrm{MW}$ (the heating power corresponds to the power flowing to the boundary region). Since the latter is a function of density and temperature profiles, its value may vary across the simulation scan. No ion temperature source is considered. The functional form of the source terms in Eqs.~\eqref{eqn:den_source}--\eqref{eqn:temp_source} is similar to that considered in the tokamak simulations of \citet{giacomin2022turbulent}.
Following the initial transient phase, the simulations reach a quasi-steady state where sources and losses balance each other. The time averaging and the statistical analysis are carried out over the final $\Delta t = 10\ R_0/c_{s0}$ interval in all the simulations. 

\subsection{Simulation results}

\begin{figure}
    \centering
    \subfloat{\includegraphics[width=\linewidth]{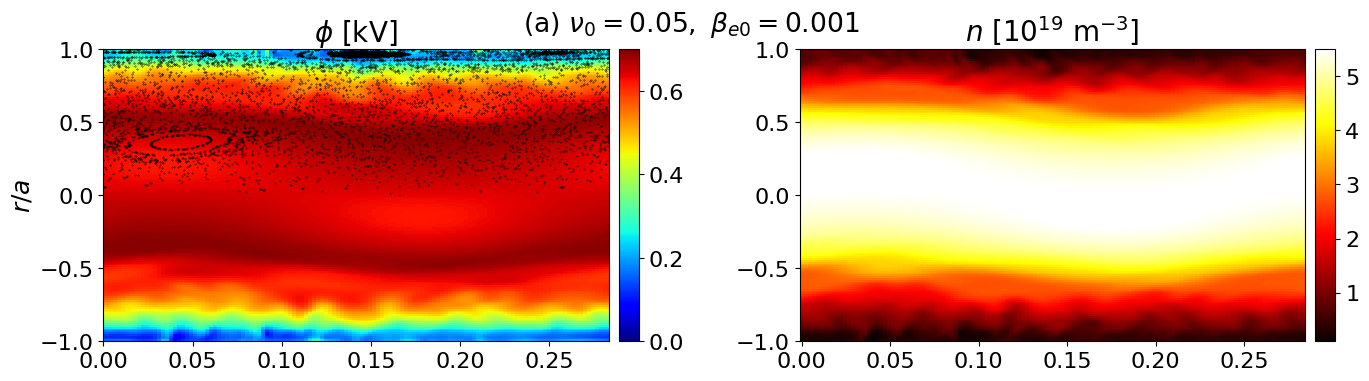}}\\
    \subfloat{\includegraphics[width=\linewidth]{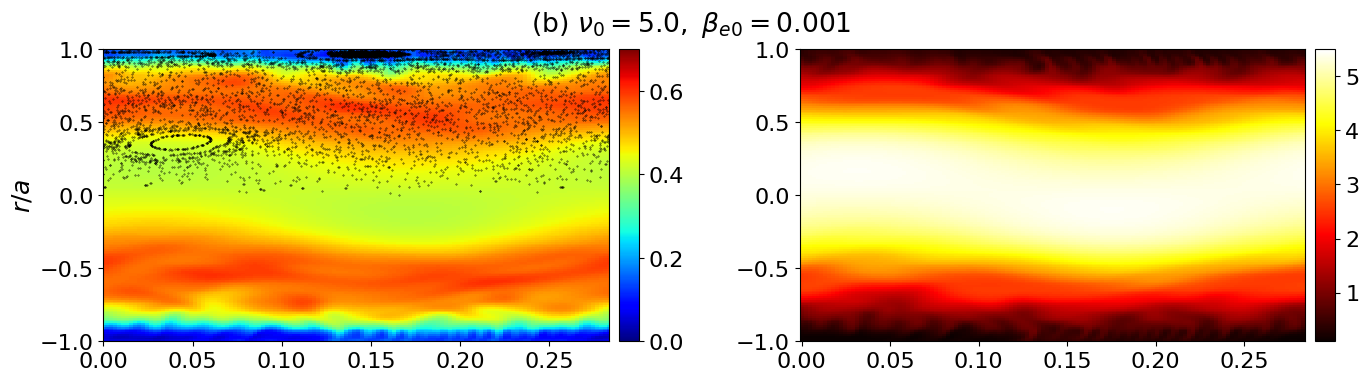}}\\
    \subfloat{\includegraphics[width=\linewidth]{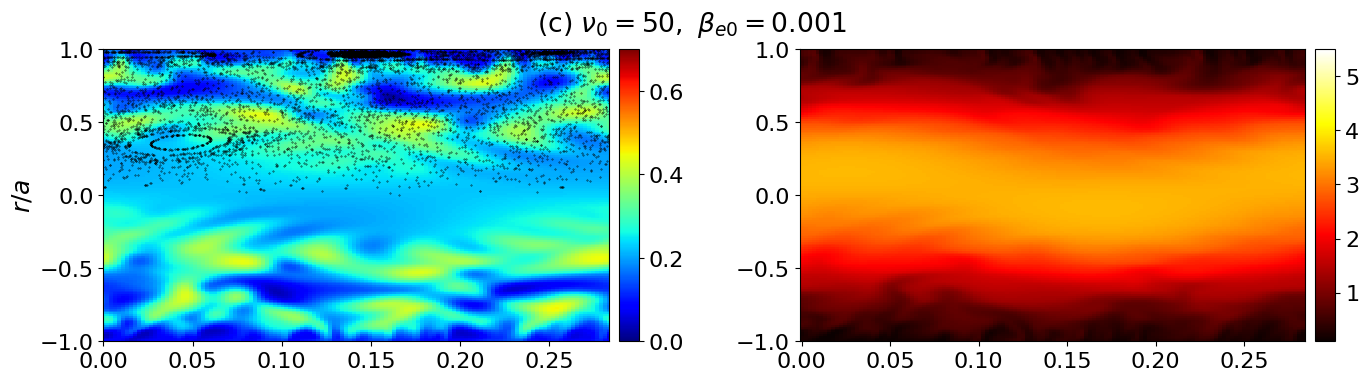}}\\
    \subfloat{\includegraphics[width=\linewidth]{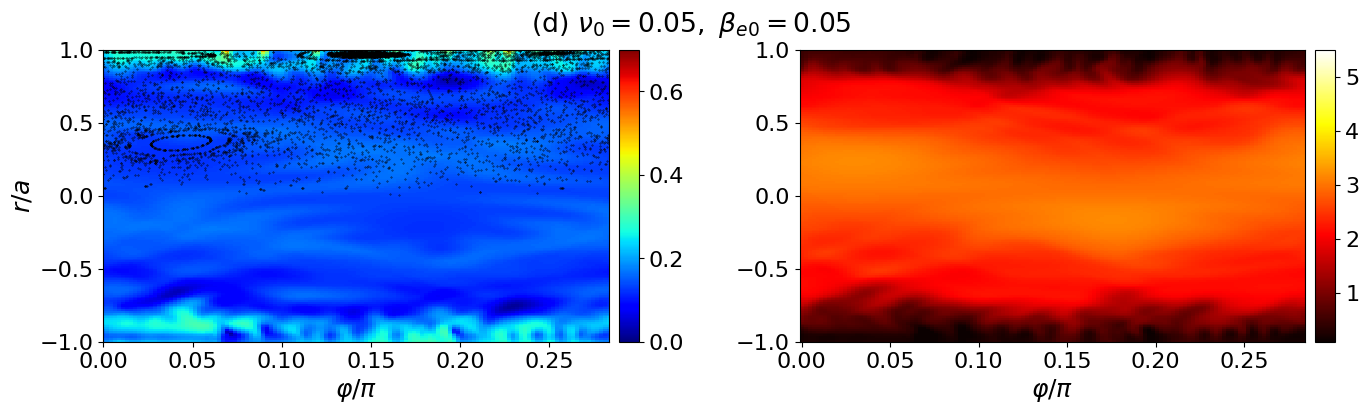}}
    \caption{Representative two-dimensional snapshots on the $R\varphi$-midplane of the electrostatic potential (left) and density (right) taken from the GBS simulations with $\nu_{0}=0.05$, $\beta_{e0}=0.001$ (a), $\nu_{0}=5.0$, $\beta_{e0}=0.001$ (b), $\nu_{0}=50.0$, $\beta_{e0}=0.001$ (c) and $\nu_{0}=0.05$, $\beta_{e0}=0.05$ (d). The Poincaré map of the magnetic field is depicted with black dots in the outer midplane region of the electrostatic potential snapshots.}
    \label{fig:snapshot}
\end{figure}

Representative two-dimensional snapshots of the density and electrostatic potential  on the $R\varphi$-midplane, obtained from four different GBS simulations characterized by $(\nu_0, \beta_{e0}) \in \{(0.05, 0.001), (5.0, 0.001), (50, 0.001), (0.05, 0.05)\}$, are presented in figure~\ref{fig:snapshot}. 
In the vicinity of the reversal surface, the $R\varphi$-midplane of RFP plasmas is almost perpendicular to the magnetic field and thus plays a role analogous to that of the poloidal plane in tokamak plasmas. 
Turbulence filaments develop in the outermost plasma region in proximity of the reversal surface. Turbulence structures observed on the high-field side ($r<0$) and on low-field side ($r>0$) are similar. This is expected since the projection of the curvature vector onto the minor radius direction in RFP plasmas is always positive, i.e., there is no good curvature region as in tokamaks.
Although the magnetic flux surfaces are disrupted, the electrostatic potential and density still conform to the structure of the underlying helical magnetic field linked to the dominant $m=1,\, n=7$ resonant mode (see the reminiscent magnetic island around $r/a\simeq 0.4$ in figure~\ref{fig:snapshot}). 
The GBS simulations indicate that the size and propagation of turbulent eddies around the $m=0, n=7$ magnetic island are essentially independent of the toroidal angle. In other words, they look the same whether they are located near the O-point or the X-point of the island. On the other hand, the electron temperature and density radial profiles are rather flat across the island. 
As a result, turbulence is only weakly driven inside the magnetic island, and the majority of turbulent structures are generated outside the island and merely pass through it.
Nonetheless, this result is in qualitative agreement with experimental observations from the thermal helium beam diagnostic~\citep{agostini2014}, where no clear difference is observed between filaments around the O-point and the X-point of the boundary magnetic island $(m, n)=(0, 7)$. 

At $\nu_0=0.05$, $\beta_{e0}=0.001$ (reference case), turbulent eddies are relatively small and limited to the region $r/a\gtrsim 0.8$ (and the symmetric region $r/a\lesssim -0.8$). 
The magnitude of turbulent fluctuation levels in the plasma core is negligible.  The electrostatic potential follows the ambipolarity response to stochastic transport, which generates an outward radial electric field proportional to the radial gradient of the electron temperature and density, as discussed in \citet{giacomin2026}.
The toroidal size of the turbulent eddies is larger at $\nu_0=5.0$, and turbulence spreads over a slightly wider radial region. Despite the two-orders-of-magnitude increase in resistivity, the two-dimensional density profile remains essentially unchanged relative to the reference case.
The turbulent transport enhancement, limited to the region $r/a>0.8$, is insufficient to cause a significant change on density.
On the other hand, the ambipolarity response of the radial electric field discussed in \citet{giacomin2026} is reduced by the increased resistivity through the term $j_\parallel/\sigma_\parallel$ in Eq.~\eqref{eqn:electron_velocity}, leading to a weaker radial electric field.
At $\nu_0=50.0$, turbulent eddies extend well into the core region, yielding a large increase of turbulent transport, which degrades the overall plasma confinement.  
This state of very large turbulent transport that appears at large collisionality shares similarities with the degraded confinement regime found in tokamak simulations~\citep{giacomin2020transp}. 
However, this regime is attained here only at substantially larger values of $\nu_0$ than those typically required in tokamak simulations (see section~\ref{sec:analysis} for a detailed discussion).
Large-scale turbulent fluctuations appear also at large values of $\beta_{e0}$. In this case, turbulence affects the entire plasma region. This regime is similar to the high-$\beta$ tokamak boundary turbulence regime identified in \citet{giacomin2022turbulent}, which corresponds to the crossing of the ideal ballooning instability boundary, but substantially larger values of $\beta_{e0}$ are required in RFP boundary turbulence simulations to attain this turbulence regime. We note that the electrostatic potential is significantly modified at large $\nu_0$ and large $\beta_{e0}$ values. This behavior arises because turbulent transport exceeds stochastic transport in the confined plasma region at sufficiently large values of $\nu_0$ and $\beta_{e0}$, thereby reducing the ambipolarity response of the electric field.  

\begin{figure}
    \centering
    \subfloat[]{\includegraphics[width=0.49\linewidth]{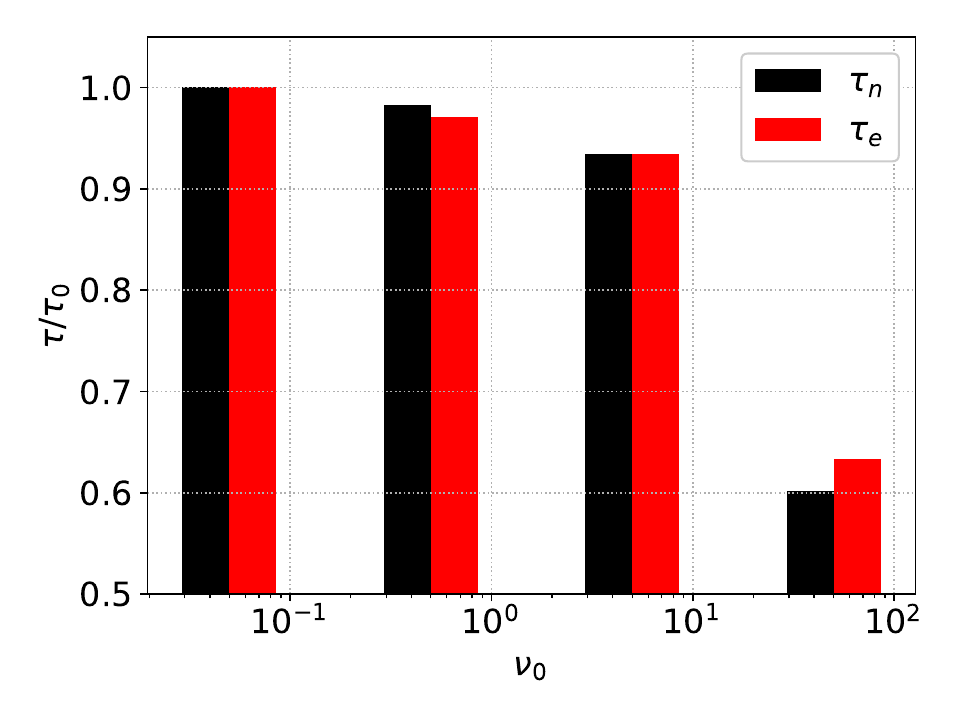}}\
    \subfloat[]{\includegraphics[width=0.49\linewidth]{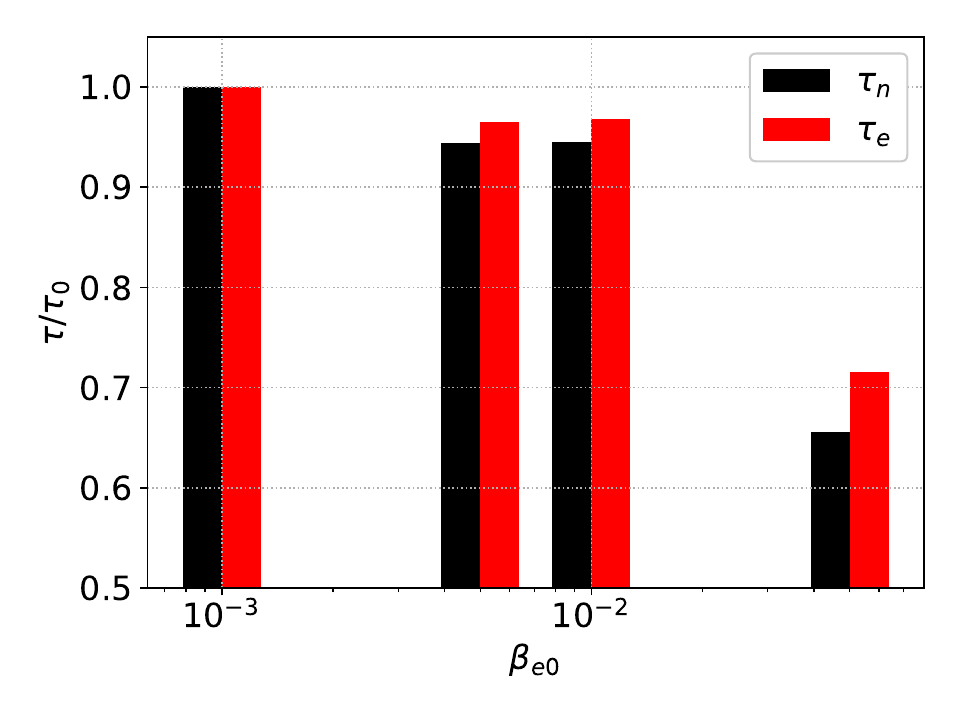}}
    \caption{Particle and energy confinement times computed from the $\nu_0$  (a) and $\beta_{e0}$ (b) simulation scans. The values are normalized to the ones computed from the reference simulation ($\nu_0 = 0.05$ and $\beta_{e0} = 0.001$).}
    \label{fig:confinement}
\end{figure}

The dependence of the particle and energy confinement times on $\nu_0$ and $\beta_{e0}$, shown in figure~\ref{fig:confinement}, supports the previous observations. The particle and energy confinement times remain nearly constant as $\nu_0$ increases from 0.05 to 5, while they drop by almost a factor of two at $\nu_0=50$. No dependence on $\beta_{e0}$ is observed until reaching $\beta_{e0} = 0.05$, where the confinement times are reduced by approximately 30\%. The trend of the particle and energy confinement times is similar. 
Notably, boundary turbulence does not affect the overall plasma confinement at intermediate and moderate values of $\nu_0$, marking a clear distinction with respect to tokamak boundary turbulence simulations~\citep{giacomin2020transp,giacomin2022turbulent}. This simulation outcome is further discussed in section~\ref{sec:analysis}.

Figure~\ref{fig:spectra} shows the $k_\varphi$ power spectrum of the density fluctuations, averaged over the radial interval $0.7 < r/a < 1$ and over time in the quasi-steady state phase, from the $\nu_0$ and $\beta_{e0}$ simulation scans. The maximum of power spectrum shifts at smaller $k_\varphi$ values as $\nu_0$ increases, although the power spectrum maximum is rather broad at $\nu_0=0.05$ and $\nu_0=0.5$. The dependence of the dominant binormal wave-vector on $\nu_0$ is consistent with resistive ballooning modes (RBMs), in agreement with \citet{giacomin2026}, which identifies RBM as the dominant instability in the reference case. 
The $\beta_{e0}$ scan reveals a weak dependence of the density power spectrum on $\beta_{e0}$, except for the case at $\beta_{e0}=0.05$, where the density power spectrum maximum shifts at the smallest finite $k_\varphi\rho_{s0}$. Turbulence in the $\beta_{e0}=0.05$ simulation is most likely affected by the reduced toroidal domain size ($L_\varphi = 2\pi/7$), which limits the minimum finite $k_\varphi\rho_{s0}$ value. 
Large-amplitude global modes develop across the full domain at large $\beta_{e0}$, with toroidal mode numbers that are comparable to the dominant low-$n$ MHD tearing modes responsible for the helical magnetic field. An accurate description of turbulence at $\beta_{e0}\gtrsim 0.05$ would likely require (i) evolving the equilibrium magnetic field, while only the perpendicular magnetic field fluctuations are typically evolved in two-fluid turbulence codes, and (ii) extending the simulation domain to the full torus. 
Nonetheless, the high-$\beta_{e0}$ simulation is consistent with the analysis at high-$\beta_{e0}$ reported in \citet{giacomin2022turbulent} for tokamak boundary turbulence simulations.

\subsection{Experimental comparison}
\begin{figure}
    \centering
    \subfloat[]{\includegraphics[width=0.49\linewidth]{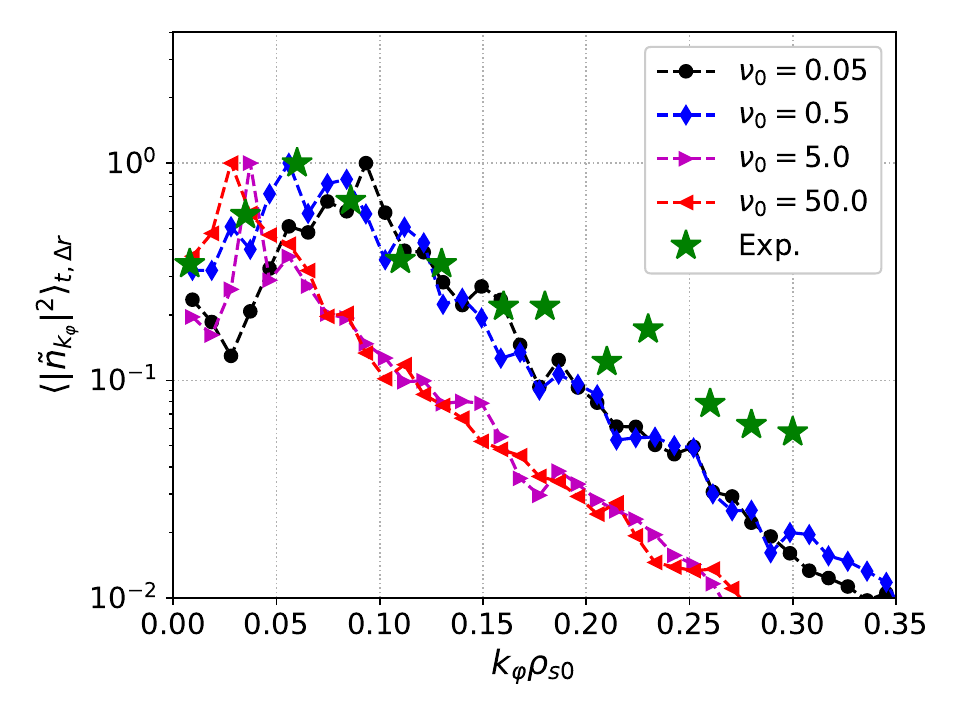}}
    \subfloat[]{\includegraphics[width=0.49\linewidth]{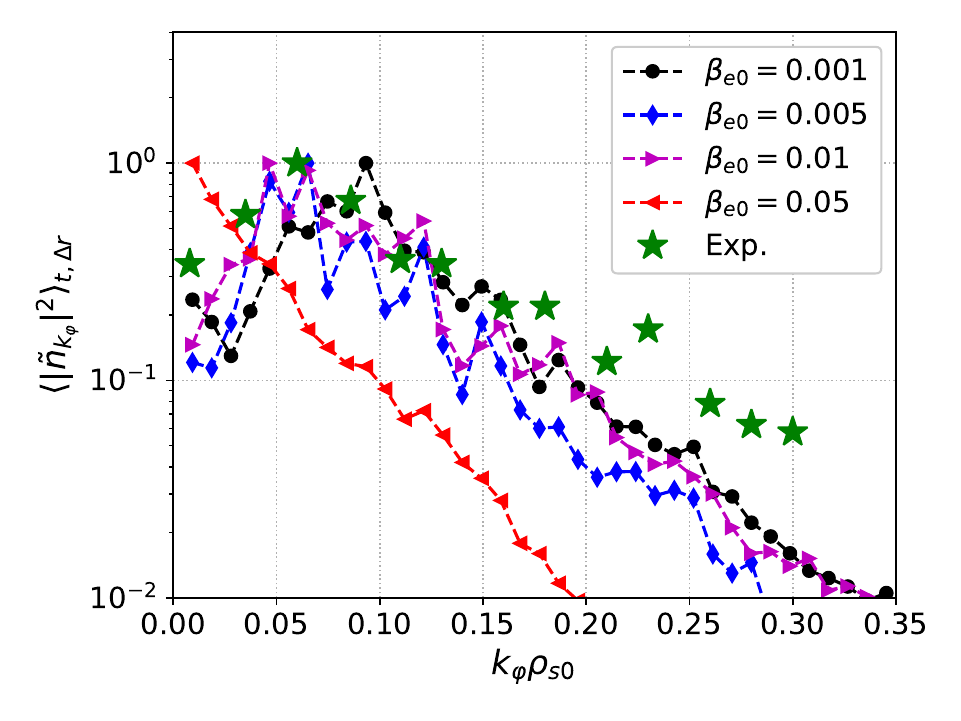}}
    \caption{Power spectrum of the density fluctuations as a function of $k_\varphi\rho_{s0}$ at various values of $\nu_0$ (a) and $\beta_{e0}$ (b), averaged over radius in the region $0.7<r/a<1$ and over time in the quasi-steady state phase. The power spectra are normalized to their maximum value. The zonal component ($k_\varphi=0$) is excluded. The experimental power spectrum, taken from \citet{rea2015}, is represented by green star-shaped markers. }
    \label{fig:spectra}
\end{figure}

The experimental power spectrum, derived from fluctuation measurements acquired by a diagnostic probe inserted into the plasma boundary region in similar conditions~\citep{spolaore2017,rea2015}, is superimposed in figure~\ref{fig:spectra}. The experimental power spectrum peaks around $k_\varphi\rho_{s0}\simeq 0.05$ and agrees fairly well with the maximum of the numerical spectrum at $\nu_0=0.5$ and $\beta_{e0}= 0.001$. The widths of the density fluctuation power spectra at $\nu_0 \simeq 0.05$ and $\nu_0 = 0.5$ are comparable to those obtained experimentally. However, the decay of the experimental power spectrum for $k_\varphi \rho_{s0} > 0.2$ is weaker than the corresponding decay observed in the numerical power spectra. This discrepancy may indicate the existence of an intermediate-$k_y$ microinstability that is not captured by the GBS simulations.

\begin{figure}
    \centering
    \subfloat[]{\includegraphics[width=0.48\linewidth]{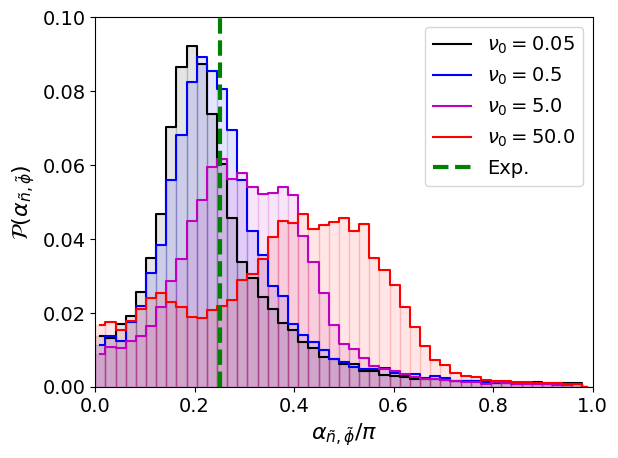}}
    \subfloat[]{\includegraphics[width=0.48\linewidth]{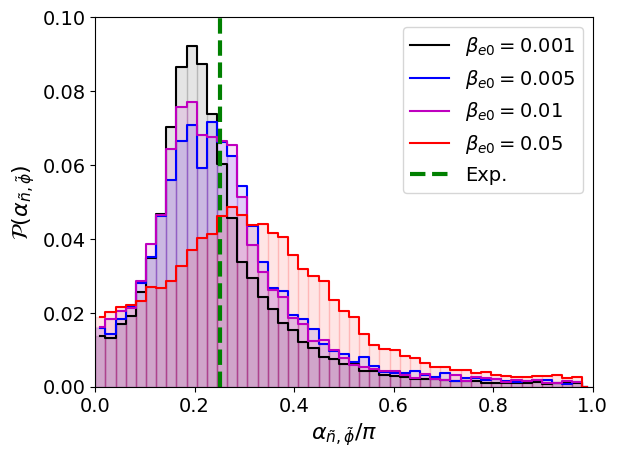}}
    \caption{Probability distribution function of the phase shift $\alpha_{\tilde{n}, \tilde{\phi}}$ between density and electrostatic potential fluctuations in the region $0.7<r/a<1$, weighted by the density fluctuation power spectrum and averaged over time in the quasi-steady state phase, at various values of $\nu_0$ (a) and $\beta_{e0}$ (b). The experimental value at $k_\varphi\rho_s\simeq 0.1$ from \citet{rea2015} is depicted as vertical green dashed line.}
    \label{fig:phase}
\end{figure}

The cross-field $\mathbf{E}\times\mathbf{B}$ particle turbulent fluxes (and analogously for the turbulent heat flux) depends on the amplitude of the density and electrostatic potential fluctuations and on their phase shift, $\Gamma_{E,r} \propto k_\varphi |\tilde{n}| |\tilde{\phi}| \sin(\alpha_{\tilde{n},\tilde{\phi}})$. At given fluctuation amplitudes, turbulent transport is maximum when $\alpha_{\tilde{n},\tilde{\phi}} = \pi/2$, which is the case for purely ballooning mode driven turbulence~\citep{Mosetto2013}.
Since the amplitude of $\tilde{n}$ and $\tilde{\phi}$ is similar across the $\nu_0$ and $\beta_{e0}$ simulation scans, the phase shift $\alpha_{\tilde{n}, \tilde{\phi}}$ provides a direct measure of the impact of turbulent transport. The distribution probability of $\alpha_{\tilde{n}, \tilde{\phi}}$ in the region $0.7 < r/a < 1$, averaged over time, is shown in figure~\ref{fig:phase} at various values of $\nu_0$ and $\beta_{e0}$. 
Starting with the $\nu_0$ scan, the most probable value of $\alpha_{\tilde{n},\tilde{\phi}}$ in the simulations with $\nu_0=0.05$ and $\nu_0=0.5$ is approximately $\pi/4$, which is in good agreement with the most probable value of $\alpha_{\tilde{n},\tilde{\phi}}$ measured experimentally~\citep{rea2015}. The $\alpha_{\tilde{n},\tilde{\phi}}$ probability distribution function broadens as $\nu_0$ increases at $\nu_0> 0.5$. At $\nu_0=50$, the $\alpha_{\tilde{n},\tilde{\phi}}$ probability distribution peaks around $\alpha_{\tilde{n},\tilde{\phi}}\simeq \pi/2$. In this case, turbulence is purely driven by RBMs. 
The parameter $\beta_{e0}$ weakly affects the probability distribution function of $\alpha_{\tilde{n}, \tilde{\phi}}$, except at $\beta_{e0}=0.05$, where it becomes broader and peaks at larger $\alpha_{\tilde{n}, \tilde{\phi}}$ values.

\begin{figure}
    \centering
    \subfloat[$\nu_0$-scan]{\includegraphics[width=0.48\linewidth]{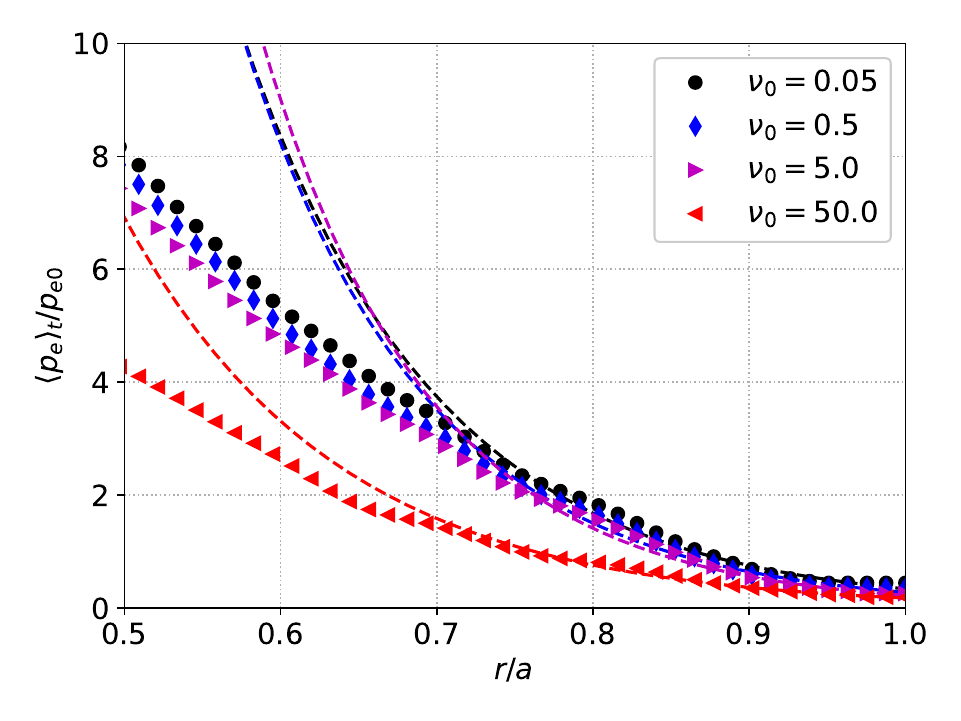}}
    \subfloat[$\beta_{e0}$-scan]{\includegraphics[width=0.48\linewidth]{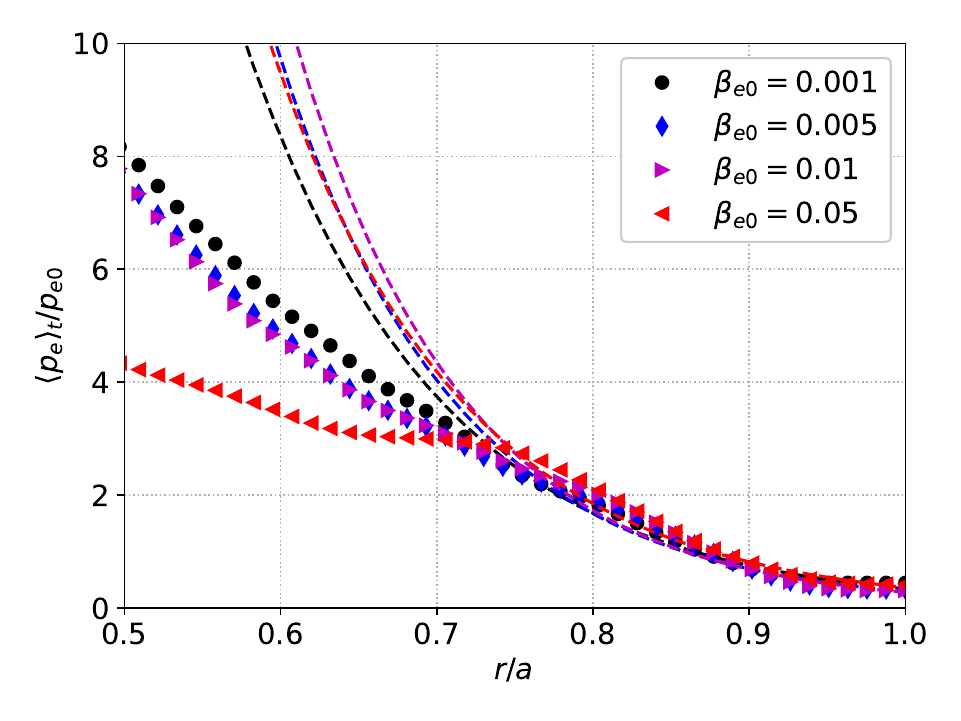}}
    \caption{Electron pressure radial profile in the outer midplane at $\varphi=0$ from the $\nu_0$ (a) and $\beta_{e0}$ (b) simulation scans, averaged over time in the quasi-steady state phase. The dashed lines represent an exponential fit of the electron pressure profile in the region $0.7<r/a<1$.}
    \label{fig:profile}
\end{figure}

Boundary turbulence emerging from this set of RFP turbulence simulations exhibits properties analogous to those typically observed in tokamak boundary turbulence simulations. This is consistent with previous experimental studies that reported similar statistical characteristics of turbulence in the boundary region of RFP and tokamak plasmas~\citep{vianello2016}.   
On the other hand, turbulent transport in RFP simulations is localized in the proximity of the reversal surface and it impacts very weakly the overall plasma confinement. 
Figure~\ref{fig:profile} displays the radial electron pressure profile at the outboard midplane obtained at $\varphi=0$ from the $\nu_0$ and $\beta_{e0}$ simulation scans, averaged over time in the quasi-steady state phase. 
The pressure profile is rather flat at $r/a\gtrsim 0.9$ at all values of $\nu_0$ and $\beta_{e0}$, whereas the overall confinement is mostly set by the pressure gradient at $r/a < 0.8$.
Since turbulence develops only over a narrow plasma region $r/a\gtrsim 0.8$, cross-field transport in the outer plasma core is mainly driven by other mechanisms, including the stochastic transport from magnetic chaos, which is independent of $\nu_0$, thus explaining the weak dependence of the radial electron pressure profile on $\nu_0$ at small and intermediate $\nu_0$ values. At $\nu_0=50$, turbulence strongly develops over a wide radial region. The enhanced turbulent transport causes a collapse of the pressure profile and the energy and particle confinement times drop substantially (see figure~\ref{fig:confinement}).
This phenomenology closely resembles the behavior observed in the vicinity of the tokamak density limit, as characterized in \citet{giacomin2022density}.

The $\beta_{e0}$ simulation scan reveals a weak dependence of the edge pressure profile on $\beta_{e0}$ at $r/a>0.7$, even at large values of $\beta_{e0}$.
In contrast, a significant flattening of the pressure profile occurs at $r/a < 0.7$ in the simulation with $\beta_{e0}=0.05$. This flattening is associated with the onset of ideal ballooning modes in the plasma core, where $\beta_e\propto p_e$ reaches values of the order of 20\%. At these large values of $\beta_{e0}$, the magnetic flutter becomes important, enhancing the stochastic transport caused by the underlying magnetic chaos.  A test simulation at even higher $\beta_{e0}$ shows destabilization of ideal ballooning modes over the whole plasma volume, including the boundary region. Subsequently, the entire pressure profile is flattened. However, describing the plasma dynamics at these large $\beta_{e0}$ values would required evolving the full equilibrium magnetic field, as done in MHD simulations, and not just the magnetic field perturbation, as typically done in two-fluid turbulence codes.  

In summary, while the RFP turbulence simulations presented here show properties similar to those typically observed in tokamak boundary turbulence simulations, they clearly point out a secondary role played by boundary turbulence on plasma confinement, marking a clear distinction with respect to tokamak boundary turbulence simulations carried out in the same geometry~\citep{giacomin2025}. This aspect is further investigated in the following section. 

\section{Turbulence regimes in the RFP plasma boundary}
\label{sec:analysis}

The theoretical framework derived in \citet{giacomin2022turbulent} links boundary turbulent transport to the edge pressure gradient and, subsequently, to the overall plasma confinement. By leveraging the set of GBS simulations presented in the previous section, this framework is revisited here for RFP plasmas.

\subsection{Turbulence phase-space in the RFP plasma boundary}
\label{sec:phasespace}

The key quantity at the base of the theoretical framework developed in \citet{giacomin2022turbulent} is the equilibrium pressure gradient length $L_p$ across the separatrix. An analytical scaling of $L_p$ has been derived in \citet{giacomin2021} by balancing the cross-field turbulent heat flux, the heat source and the parallel losses to the target plates, assuming (i) turbulence  to be driven by RBMs and (ii) parallel transport in the scrape-off layer (SOL) to be mostly convective, leading to\footnote{We note that Eq.~\eqref{eqn:lp_tok} has been extended in \cite{lim2023} to account for plasma shaping effects. In this work, we use Eq.~\eqref{eqn:lp_tok} since RFP plasmas have mostly a circular poloidal cross section with negligible shaping.}
\begin{equation}
\label{eqn:lp_tok}
    L_p \simeq \frac{5^{8/17}\pi^{12/17}}{2^{13/17}}\biggl(\rho_*^3\nu_0^6 q^{12}a^{12}(1+\kappa^2)^6n^{10}S_p^{-4}\biggr)^{1/17}\,,
\end{equation}
where $L_p$ is normalized to $\rho_{s0}$, $q$ is the edge safety factor, $a$ is the plasma minor radius normalized to $\rho_{s0}$, $\kappa$ is the edge elongation, $n$ is the normalized separatrix density and $S_p= \rho_* \int s_p \mathrm{d}R\mathrm{d}Z$ is the normalized power source. Unless specified otherwise, dimensionless quantities are used in the following.

Eq.~\eqref{eqn:lp_tok} assumes that the binormal size of RBMs is limited by the parallel current, i.e., the curvature and the parallel gradient terms in Eq.~\eqref{eqn:vorticity} balance each others~\citep{giacomin2020transp}, $2\mathcal{C}(\tilde{p}_e) \sim \nabla_\parallel \tilde{j}_\parallel$. The pressure fluctuation amplitude, $\tilde{p}_e$, is obtained from the linearized electron pressure equation,  $\tilde{p}_e\simeq i k_y \tilde{\phi}p_e/(\rho_*L_p)$, while the electrostatic potential fluctuation amplitude, $\tilde{\phi}$, is estimated from the linearized Ohm's law, $i k_\parallel \tilde{\phi}\sim \nu \tilde{j}_\parallel$, where $k_y$ is the binormal wave-vector (normalized to $1/\rho_{s0}$), $k_\parallel$ is the parallel wave-vector (normalized to $1/R_0$) and $\gamma=\sqrt{2T_e/(\rho_*L_p)}$ is the maximum growth rate of the interchange instability.
Therefore, $k_y$ is simply approximated by $k_y \sim k_\parallel /\sqrt{n\nu \gamma}$.
The parallel connection length of an axisymmetric toroidal plasma is defined as $L_\parallel = 2\pi R\sqrt{q^2+\epsilon^2}$, with $\epsilon=r/R$. Within the boundary region of a tokamak, $L_\parallel$ reduces to $L_\parallel^\mathrm{tok} \simeq 2\pi R_0 q$, which yields $k_\parallel \simeq 1/q$ and, consequently, leads to Eq.~\eqref{eqn:lp_tok}. 
In contrast, in the boundary region of an axisymmetric RFP plasma, where $q \simeq 0$, $L_\parallel$ can be approximated as $L_\parallel^\mathrm{RFP} \simeq 2\pi a$.

The GBS simulations presented in section~\ref{sec:simulations} show that turbulence is mainly driven by RBMs. Therefore, we might expect the quasi-linear theory derived in \citet{giacomin2020transp} to hold in the boundary of RFP plasmas as long as (i) $k_\parallel$ is correctly evaluated and (ii) turbulence provides the dominant cross-field transport channel. By making the $k_\parallel$ dependence explicit in Eq.~\eqref{eqn:lp_tok}, we obtain
\begin{equation}
\label{eqn:lp_rfp}
    L_p \simeq 2^{5/17} 5^{8/17}\pi^{12/17}\biggl(\rho_*^3\nu_0^6 k_\parallel^{-12}a^{12} n^{10}S_p^{-4}\biggr)^{1/17}\,,
\end{equation}
which is identical to Eq.~\eqref{eqn:lp_tok} except for the dependence on the safety factor and on the elongation, which is taken as $\kappa=1$ (RFP plasmas are typically circular).
The difference on the upfront numerical factor is partially due to $\kappa=1$ in Eq.~\eqref{eqn:lp_rfp} and partially comes from considering turbulent transport uniform on the poloidal angle, while, in \citet{giacomin2020transp}, the turbulent flux is assumed to vanish on the high-field side, i.e., in the good curvature region. 
By neglecting the scattering of magnetic field lines due to magnetic chaos, $k_\parallel$ in the proximity of the reversal surface can be approximated by $k_\parallel\simeq 1/(\rho_*a)$. This can be considered as an upper limit for $k_\parallel$, since magnetic chaos and the helical structures breaking axisymmetry are likely to increase the parallel connection length (see section~\ref{sec:limits} for a discussion on the parallel connection length). 
We note that the primary difference between Eq.~\eqref{eqn:lp_rfp} and the $L_p$ scaling derived in \cite{giacomin2020transp} concerns $k_\parallel$, whereas the effects of toroidal shaping associated with $n=7$, $m \in \mathbb{N}$ tearing modes, as well as the resulting magnetic chaos, are neglected.

Three turbulence regimes have been identified in \citet{giacomin2022turbulent} depending on the value of plasma collisionality and $\beta_{e0}$: a regime at low collisionality, where turbulence is driven by drift-waves, a regime at intermediate and large collisionality, where turbulence is driven by RBMs, and a regime at high $\beta_{e0}$ where turbulence is driven by ideal ballooning modes. In the RBM regime, turbulent transport increases with the edge collisionality, leading to a collapse of the pressure profile at large density, characterized by $L_p\sim a$. This condition has been identified as the crossing of a density limit~\citep{giacomin2022density}. Strongly enhanced turbulent transport and the subsequent flattening of the edge pressure profile are observed in the large $\nu_0$ simulation in section~\ref{sec:simulations}, thus suggesting a possible similar phenomenology for RFP plasmas. 
Hence, this limit is revisited here using the equilibrium pressure gradient length scaling in Eq.~\eqref{eqn:lp_rfp}.
The condition $L_p\sim a$ leads to 
\begin{equation}
\label{eqn:den_lim}
    \frac{\nu_0^{3/2}}{S_p} \sim \frac{1}{2^{5/4}25\pi^3}\frac{a^{5/4}k_\parallel^3}{\rho_*^{3/4}n^{5/2}}\,.
\end{equation}
Except for the upfront numerical factor, Eq.~\eqref{eqn:den_lim} is equivalent to Eq.~(45) of \citet{giacomin2022turbulent} if $k_\parallel=1/q$ and $\kappa=1$.
We emphasize that $\nu_0^{3/2}/S_p$ at the transition depends strongly on $k_\parallel$.
The corresponding density limit condition for RFP plasmas is obtained by substituting $k_\parallel = 1/(\rho_*a)$ in Eq.~\eqref{eqn:den_lim}, leading to
\begin{equation}
    \label{eqn:den_lim_rfp}
    \frac{\nu_0^{3/2}}{S_p} \sim \frac{1}{2^{5/4}25\pi^3}\frac{1}{\rho_*^{15/4}n^{5/2}a^{7/4}}\,.
\end{equation}

\begin{figure}
\centering
    \subfloat[]{\includegraphics[width=0.49\linewidth]{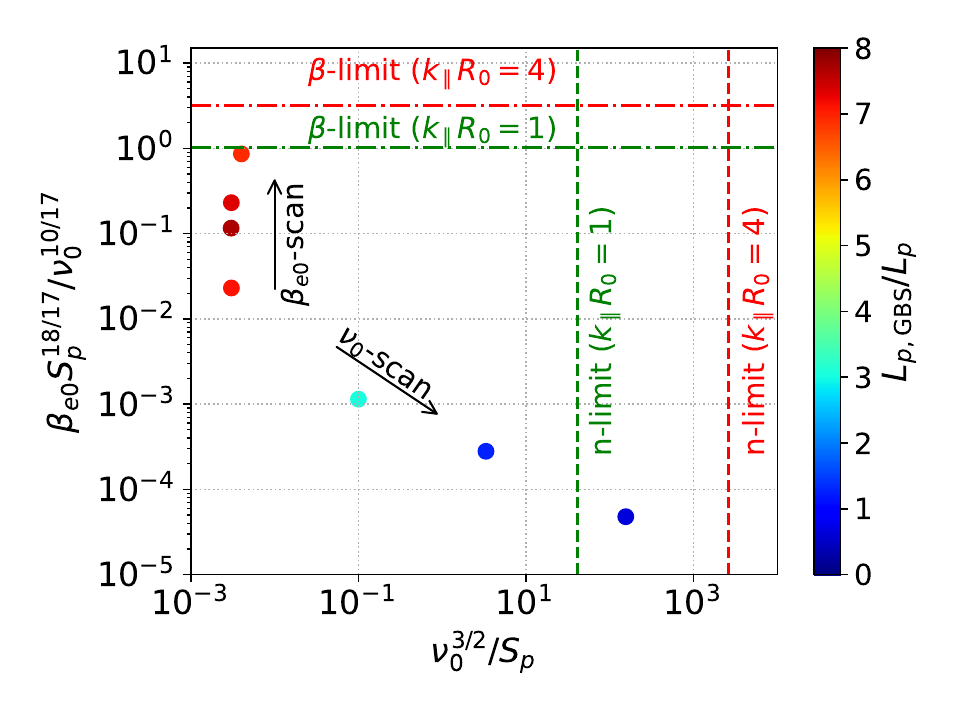}}
    \subfloat[]{\includegraphics[width=0.49\linewidth]{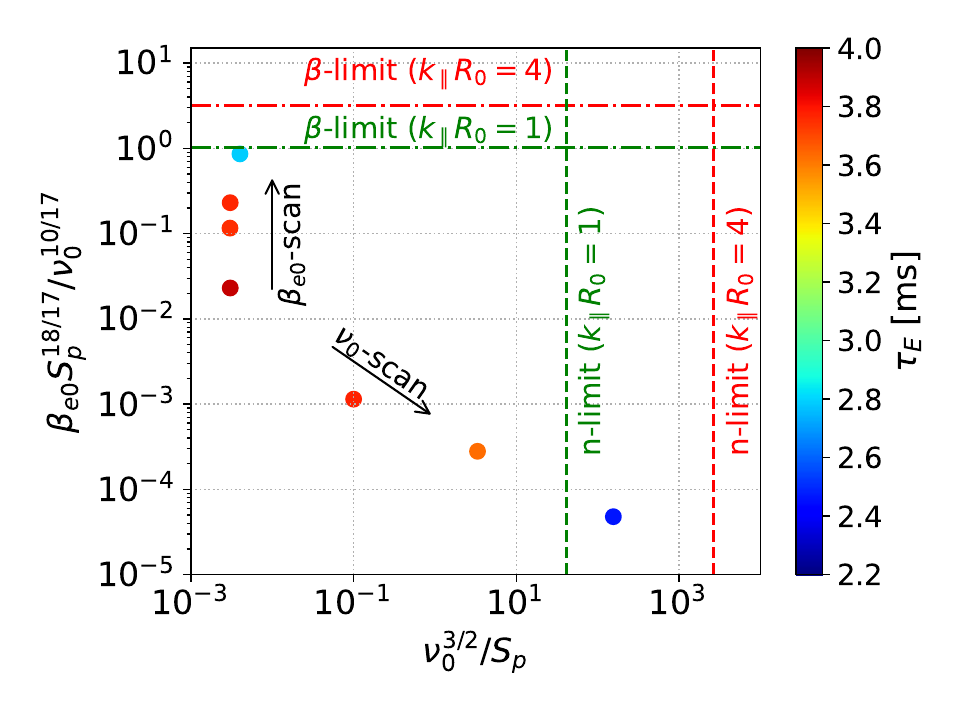}}
    \caption{Phase space of RFP boundary turbulence. Each dot in the figures corresponds to a GBS simulation, while the $\beta_{e0}$ and $\nu_0$ scans are identified with a label. The colorscale represents the ratio of the numerical to analytical edge pressure gradient length (a) and the energy confinement time (b). The vertical and horizontal dashed lines indicate the limits predicted by Eq.~\eqref{eqn:den_lim} and Eq.~\eqref{eqn:beta_limit}, respectively, for two different values of $k_\parallel$. }
    \label{fig:phase_space}
\end{figure}

Figure~\ref{fig:phase_space}(a) illustrates the dependence of the ratio between the numerically evaluated pressure gradient scale length, obtained from GBS simulations by fitting the radial profiles of the electron pressure in the interval $0.7 < r/a < 1$ (see figure~\ref{fig:profile}), and the analytical prediction given by Eq.~\eqref{eqn:lp_rfp}, as a function of the parameter $\nu_0^{3/2}/S_p$. At low $\nu_0^{3/2}/S_p$, the numerical $L_p$ exceeds significantly the value predicted by Eq.~\eqref{eqn:lp_rfp}, while the ratio $L_{p,\mathrm{GBS}}/L_p$ approaches unity at large $\nu_0^{3/2}/S_p$. This is consistent with RBM-driven turbulent transport becoming increasingly important in the region $0.7 < r/a < 1$ as $\nu_0$ increases. In other words, the $L_p$ scaling in Eq.~\eqref{eqn:lp_rfp} strictly applies only at $\nu_0^{3/2}/S_p>0.1$, when RBM turbulence is the dominant cross-field transport channel. It is worth noting that, because of the large $k_\parallel$ value, Eq.~\eqref{eqn:lp_rfp} predicts a very steep pressure gradient at low $\nu_0$, i.e., the pressure profile would be much steeper if exclusively limited by RBM-driven turbulent transport. 
Analogously, figure~\ref{fig:phase_space}(b) shows the energy confinement time reduction associated with the increase of the RBM-driven turbulent transport. The energy confinement time drops by a factor of two before crossing the limit predicted by Eq.~\eqref{eqn:den_lim_rfp}, which is derived from Eq.~\eqref{eqn:den_lim} by assuming $k_\parallel = 1/(\rho_*a)$ (corresponding to $k_\parallel R_0 = R_0/a=4$ for a RFX-mod plasma). This is expected as Eq.~\eqref{eqn:den_lim_rfp} provides only an upper limit. The value of $k_\parallel=2\pi/L_\parallel$ might be smaller than that estimated from magnetic field geometry considerations (see the discussion in section~\ref{sec:limits}). For instance, the limit occurs at much smaller $\nu_0^{3/2}/S_p$ values with $k_\parallel =1 $, and the simulation with $\nu_0=50$ would be above this limit. 

The phase space of boundary turbulence in figure~\ref{fig:phase_space} retrieves the connection between boundary turbulence and energy confinement previously found in tokamak turbulence simulations~\citep{giacomin2022turbulent}.
On the other hand, the density limit in figure~\ref{fig:phase_space} is crossed only at unrealistically large values of $\nu_0$, even when $k_\parallel = 1$. In fact, the density limit previously derived for tokamak plasmas is approximately two orders of magnitude smaller than that predicted by Eq.~\eqref{eqn:den_lim_rfp},
\begin{equation}
    \frac{(\nu_0^{3/2}/S_p)_\mathrm{RFP}}{(\nu_0^{3/2}/S_p)_\mathrm{tokamak}} \sim \Bigl(\frac{q}{\rho_*a}\Bigr) \gtrsim 100\,.
\end{equation}
This result clearly questions the applicability of this theoretical framework in the edge region of RFP plasmas, where transport mechanisms other than RBM-driven turbulence appear to control the edge pressure gradient at realistic values of $\nu_0$.   

The phase space of tokamak edge turbulence of \citet{giacomin2022turbulent} provides also a limit for the maximum achievable value of $\beta_{e0}$ before a substantial reduction of the energy confinement time is observed in tokamak turbulence simulations. This limit was derived from the parameter $\alpha_\mathrm{MHD} = -R_0 q^2 \mathrm{d}\beta/\mathrm{d}r$, which is ill-defined in the boundary of RFP plasmas where $q$ vanishes. Alternatively, this limit can be derived directly from the condition corresponding to the onset of ideal ballooning modes, which is obtained from the generalized Ohm's law (see Eq.~\eqref{eqn:electron_velocity}), by balancing the electromagnetic induction and the resistive current terms,
\begin{equation}
    \label{eqn:ohm}
    \gamma \psi \sim \nu \tilde{j}_\parallel\,,
\end{equation}
where $\gamma$ is the maximum growth rate of the ballooning instability, $\gamma = \sqrt{2T_e/(\rho_* L_p)}$. By using Eq.~\eqref{eqn:ampere}, the magnetic fluctuation $\psi$ can be approximated by $\psi \sim \beta_{e0} j_\parallel /(2 k_\perp^2)$, where $k_\perp$ is the perpendicular wave vector. In the boundary region of RFP plasmas, $k_\perp$ can be replaced by $k_\varphi$. Furthermore, before the onset of ideal ballooning modes, turbulence is driven by RBMs and $k_\varphi^2$ can be approximated by $k_\varphi^2 \simeq k_\parallel^2/(n\nu\gamma)$.      
Within these assumptions, Eq.~\eqref{eqn:ohm} leads to $\beta_{e0}\sim k_\parallel^2\rho_* L_p/p_e$. The $\beta$ limit is finally obtained by replacing $L_p$ with Eq.~\eqref{eqn:lp_rfp},
\begin{equation}
    \label{eqn:beta_limit}
    \frac{\beta_{e0}S_p^{18/17}}{\nu_0^{10/17}} \sim 2^{3/17}5^{2/17}\pi^{55/51}\rho_*^{22/17}n^{11/17}k_\parallel^{14/17}a^{20/17}\,,
\end{equation}
where the key parameters are isolated in the left-hand side of Eq.~\eqref{eqn:beta_limit}. Except for the numerical pre-factor, Eq.~\eqref{eqn:beta_limit} is identical to Eq.~(50) of \citet{giacomin2022turbulent} if $k_\parallel = 1/q$ and $\kappa = 1$. 

As discussed in section~\ref{sec:simulations}, turbulent transport in the region $0.7<r/a<1$ is weakly affected by the parameter $\beta_{e0}$, and the GBS simulations belonging to the $\beta_{e0}$ scan return similar values of $L_{p, \mathrm{GBS}}$ (see figure~\ref{fig:profile}). Consequently, figure~\ref{fig:phase_space}(a) shows that $L_{p, \mathrm{GBS}}/L_p$ is nearly independent from $\beta_{e0}$ and it largely exceeds unity.
Moreover, figure~\ref{fig:phase_space}(b) shows that the energy confinement time drops as the $\beta$ limit, given by Eq.~\eqref{eqn:beta_limit}, is approached, providing support to the applicability of Eq.~\eqref{eqn:beta_limit}. 
We note that, the $\beta$ limit predicted by Eq.~\eqref{eqn:beta_limit} is less sensitive on the $k_\parallel$ assumption than the density limit predicted by Eq.~\eqref{eqn:den_lim}.

\subsection{Limits of the two-fluid model approach in RFP}
\label{sec:limits}

The statistical properties of turbulence fluctuations in the plasma boundary of the GBS simulations presented in section~\ref{sec:simulations} compare reasonably well to experiments. Although a more comprehensive validation would require full-scale GBS simulations (at least in the poloidal plane), this comparison provides supporting evidence for the reliability of the turbulence properties predicted by GBS simulations. 
RBM-driven turbulence is also observed in RFX-mod GBS tokamak simulations~\citep{giacomin2025}, thus proving a strong connection between RFP and tokamak boundary turbulence, which supports the experimental finding of \citet{vianello2016}.
Furthermore, the turbulence phase-space developed in section~\ref{sec:phasespace} for RFP GBS turbulence simulations shares the same theoretical framework as the one developed for tokamak edge turbulence~\citep{giacomin2022turbulent}.
These results underscore the significant insights that two-fluid, global, flux-driven turbulence simulations can offer for disentangling the complex turbulent dynamics operating at the edge region of RFP plasmas.  
However, there are a few criticalities. First, the equilibrium profiles and the energy confinement depend weakly on $\nu_0$ and $\beta_{e0}$ at realistic values of $\nu_0$ and $\beta_{e0}$, while these are key parameters controlling tokamak boundary turbulence.
Second, turbulence develops only in a very narrow region at the plasma boundary unless unrealistically large values of $\nu_0$ and $\beta_{e0}$ are considered. 
Finally, the crossing of the density limit predicted by Eq.~\eqref{eqn:den_lim_rfp} occurs at unrealistically large values of density.  

The fluid closure used in Braginskii models requires the electron mean free path, $\lambda_e = v_{th,e}\tau_e$ ($v_{th,e} =\sqrt{T_e/m_e}$ is the thermal electron velocity and $\tau_e = 0.51 m_e \sigma_\parallel/(ne^2)\propto \nu^{-1}$ is the electron collision time), to be smaller than the typical length scale in the parallel direction, known as parallel connection length, $L_\parallel$. This condition is typically satisfied in present-day tokamak devices, although kinetic effects, which are not captured by a collisional closure, may still play an important role~\citep{oliveira2022,zholobenko2021,pitzal2023,frei2025}.

The definition of a parallel connection length in RFP plasmas is nontrivial. In the vicinity of the reversal surface, simple magnetic field geometry considerations would indicate a $L_\parallel$ of $2\pi a$. A turbulent eddy localized on the reversal surface forms a closed structure after completing a single poloidal turn, therefore constraining its parallel extension to $2\pi a$. However, when moving radially inwards from the reversal surface, additional factors must be taken into account, including the helical deformation of the magnetic field and the presence of magnetic field chaos.
Another key factor that enhances ballooning modes in RFP plasmas is the lack of regions with favorable magnetic curvature. Consequently, the commonly invoked argument that ballooning modes extend along magnetic field lines until they encounter a region of good curvature, where they are stabilized, does not apply to RFP configurations. In this geometry, the parallel correlation length of RBMs can exceed one full poloidal transit without the modes being quenched on the high–field side. This feature effectively increases the role of ballooning instabilities in the plasma, thus providing some support to the GBS results.    
Nonetheless, it is reasonable to assume that $L_\parallel$ at the edge of RFP plasmas remains shorter than $L_\parallel$ in tokamak plasmas. Hence, at similar values of electron temperature and density (similar $\lambda_e$), the collisionality in RFP plasmas is expected to be smaller than the one in tokamaks.

\begin{figure}
    \centering
    \subfloat[]{\includegraphics[width=0.51\linewidth]{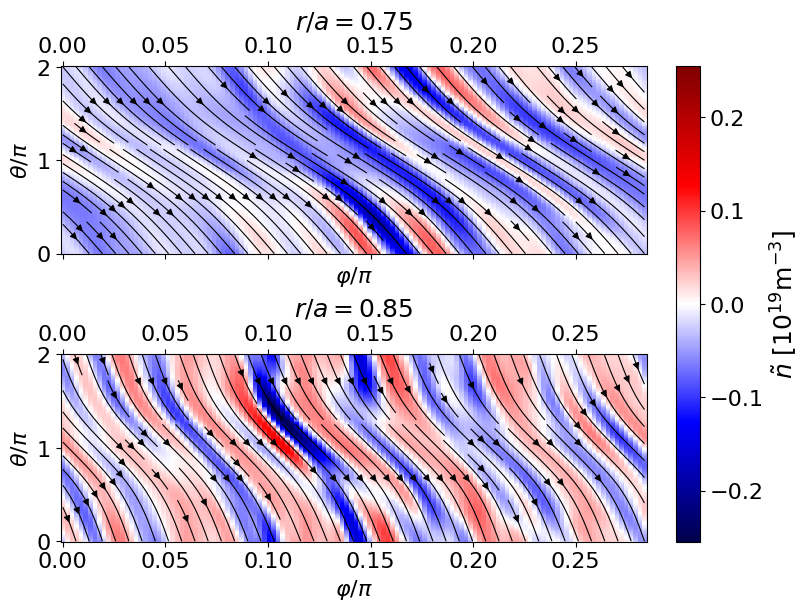}}\,
    \subfloat[]{\includegraphics[width=0.46\linewidth]{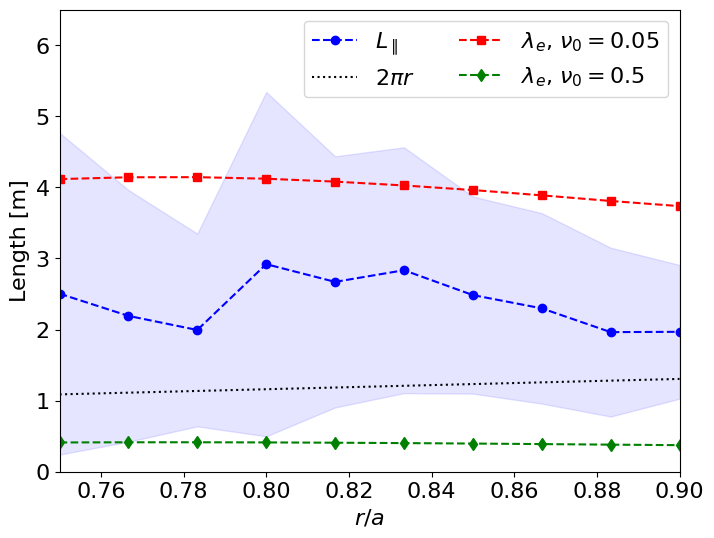}}
    \caption{(a) Two-dimensional snapshots of density fluctuations on the $(\theta, \varphi)$ surface at two different radial locations from the reference simulation. The black curves represent the magnetic field lines at the surface. (b) Comparison between the ensemble-averaged parallel correlation length and the electron mean free path obtained from simulations with $\nu_0=0.05$, $\beta_{e0} = 10^{-3}$ and with $\nu_0=0.5$, $\beta_{e0} = 10^{-3}$. The shaded region denotes the standard deviation of $L_\parallel$, while the black dotted line marks $2\pi r$.}
    \label{fig:lpar}
\end{figure}

The parallel correlation length of turbulent fluctuations can be directly obtained from the simulation results. Figure~\ref{fig:lpar}(a) presents two snapshots of $\tilde{n}$ on the $(\theta, \varphi)$ surface, taken at two distinct radial positions of the reference simulation. On both flux surfaces, the density perturbations display a characteristic parallel scale that extends beyond one poloidal turn. The parallel correlation length, which can be effectively identified with the characteristic parallel scale \(L_\parallel\), is here defined as the parallel spatial separation between two consecutive zeros of the density fluctuation field.
The ensemble average of $L_\parallel$ is displayed in figure~\ref{fig:lpar}(b) at different radial positions. The geometrical estimate $L_\parallel \simeq 2\pi r$ is also displayed in figure~\ref{fig:lpar}(b) for reference. Although $L_\parallel$ exhibits substantial scatter, the majority of the fluctuation structures possess a parallel correlation length that exceeds the value $2\pi r$ by a factor of approximately 2–3. This yields a value of $k_\parallel$ such that $1 < k_\parallel R_0 \leq 4$, thereby justifying the bounds adopted in section~\ref{sec:phasespace}.  

The surface-averaged value of $\lambda_e$ obtained from the reference simulation, together with that computed from the simulation with $\nu_0 = 0.5$ and $\beta_{e0} = 10^{-3}$, is also presented in figure~\ref{fig:lpar}(b). The electron mean free path in the reference simulation exceeds the parallel connection length across the entire radial region of figure~\ref{fig:lpar}(b), although the significant scattering affecting $L_\parallel$ yields comparable values in the region $0.8\lesssim r/a \lesssim 0.85$. Conversely, the simulation with higher $\nu_0$, i.e., higher $n_0$, yields $\lambda_e<L_\parallel$. In the latter case, $\lambda_e$ is smaller than the geometrical $L_\parallel$. 
This is expected as $\lambda_e$ is inversely proportional to $\nu_0$. Hence, the condition $L_\parallel/\lambda_e>1$ is fulfilled for values of $\nu_0$ larger than the reference one. Moreover, at $\nu_0\gtrsim 5$, $\lambda_e$ becomes significantly smaller than $L_\parallel$ over a wide radial region, thus allowing RBM turbulence to easily develop in the core plasma region and therefore controlling the overall plasma confinement. This explains why the energy confinement time is found to depend on boundary turbulence only at (unrealistically) large values of $\nu_0$. 
It is worth remarking that the $k_\parallel$ stabilization mechanism remains strong even at large $\nu_0$, in agreement with Eq.~\eqref{eqn:lp_rfp}.
Notably, the stabilization mechanism associated with finite $k_\parallel$ is sufficiently strong that the pressure gradient is primarily limited by transport channels other than RBM-induced turbulent transport, except in the regime where $\nu_0$ is increased to unrealistically large values. This occurs despite the fact that RBM turbulence does develop and produces power-density fluctuation spectra that exhibit reasonably good quantitative agreement with experimental measurements (see figure~\ref{fig:spectra}).

In conclusion, the short parallel connection length in RFP plasmas clearly restricts the validity of two-fluid turbulence simulations. This result calls for future investigations with more advanced modes, such as the one recently implemented in GRILLIX~\citep{pitzal2023}.

\section{Linear gyrokinetic analysis in the outer core}
\label{sec:gyrokinetic}

The analysis presented in section~\ref{sec:limits} highlights some of the limits behind the fluid description of plasma turbulence in the edge of RFP plasmas, calling for the need of more advanced models, such as gyrokinetic. 
Although a comprehensive gyrokinetic analysis lies beyond the scope of the present study, it is nonetheless important to examine which microinstabilities are expected to be active in the outer core region according to a gyrokinetic model. 
We perform a set of local (flux-tube) linear gyrokinetic simulations using the GENE code~\citep{jenko2000}.
The simulations are performed in outer core at a normalized minor radius of $r/a = 0.8$.
Since the helical deformation associated with the $m=1$, $n=7$ magnetic island is predominantly confined within the radial domain $r/a < 0.8$, the equilibrium magnetic configuration in this region can be reasonably modeled as circular and axisymmetric. Consequently, it is represented using a Miller parametrization~\citep{miller1998} characterized by a safety factor $q = 0.06$ and a magnetic shear $\hat{s} = -10$. 
While considering an outer surface would likely enable a more direct comparison with the two-fluid turbulence simulations, the choice of this location is strongly constrained by the presence of the $q=0$ surface and the associated $m=0$, $n=7$ magnetic islands. The $r/a=0.8$ surface is therefore an appropriate compromise, balancing the need to avoid the $m=0$, $n=7$ magnetic islands while remaining as close as possible to the plasma edge. 
Another important limitation of the magnetic equilibrium adopted in this gyrokinetic analysis is the neglect of magnetic chaos. The inclusion of magnetic chaos into gyrokinetic simulations presents a substantial methodological challenge that lies beyond the scope of the present work.
Despite these limitations, the present linear gyrokinetic analysis can still yield valuable insights into the role of kinetic instabilities in the edge of RFP plasmas and serve as a basis for guiding future more advanced gyrokinetic investigations.

The normalized local gradients of density and temperature on the $r/a=0.8$ magnetic flux surface, computed from the GBS equilibrium profiles, are $(a/n_e)\,|\partial_r n_e| = (a/n_i)\,|\partial_r n_i| = 13$, $(a/T_e)\,|\partial_r T_e| = 7.5$, and $(a/T_i)\,|\partial_r T_i| = 10$, where we set $a = 0.46\ \mathrm{m}$, which is the minor radius of RFX-mod and which is adopted as the reference length in the gyrokinetic simulations. The electron beta is set to $\beta_e = 0.001$. Both the electrostatic potential, $\delta \phi$, and the parallel vector potential, $\delta A_\parallel$, are evolved. Given the low value of $\beta_e$, parallel magnetic fluctuations are not included in the simulations.
The numerical resolution employed in the simulations is given by $(n_x, n_z, n_v, n_\mu) = (32, 64, 32, 16)$, where $n_x$ and $n_z$ denote the number of grid points in the radial and parallel spatial directions, respectively, and $n_v$ and $n_\mu$ represent the number of grid points used for the discretization of the parallel velocity and magnetic moment, respectively. A resolution convergence study is reported in appendix~\ref{app:resolution}.

\begin{figure}
\centering
    \subfloat[]{\includegraphics[width=0.49\linewidth]{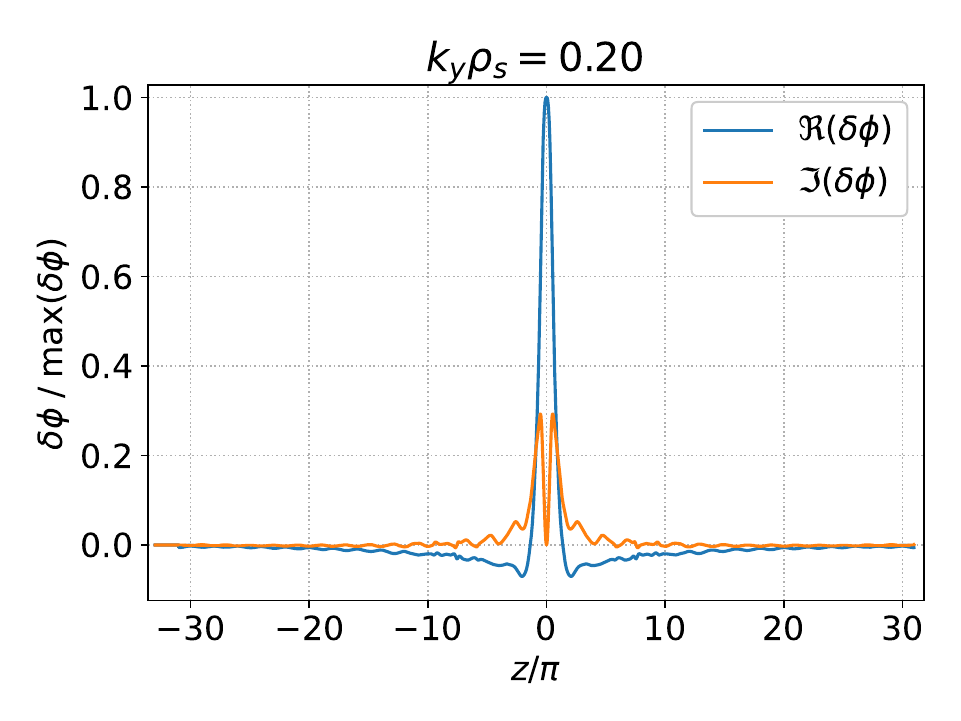}}
    \subfloat[]{\includegraphics[width=0.49\linewidth]{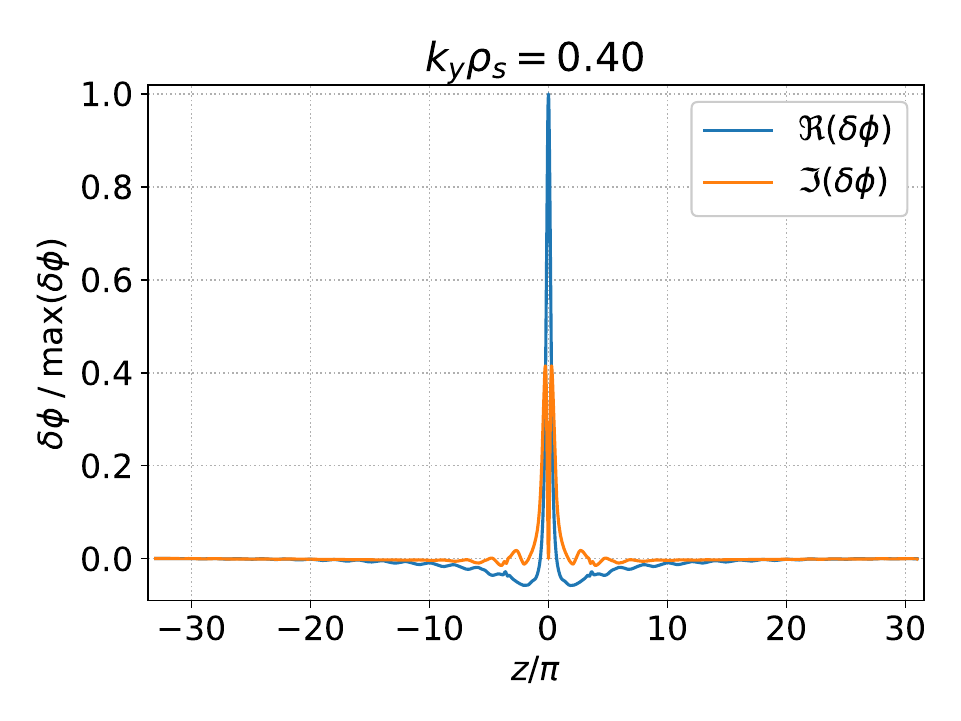}}
    \caption{Real and imaginary part of $\delta \phi$ as a function of the parallel coordinate $z$ for the $k_y\rho_s=0.2$ (a) and $k_y\rho_s=0.4$ (b) unstable modes. }
    \label{fig:field}
\end{figure}

Figure~\ref{fig:growth_rate} shows the growth rate and the real mode frequency as functions of the binormal wave-vector $k_y\rho_s$. Unstable modes are found at ion scale ($0.1 < k_y\rho_s < 0.7$) with negative mode frequency (drift velocity in the electron diamagnetic direction). No unstable mode is found at electron scale. The growth rate spectrum exhibits two local maxima at $k_y\rho_s = 0.2$ and $k_y\rho_s=0.4$. The discontinuity in the frequency spectrum between $k_y\rho_s=0.2$ and $k_y\rho_s=0.25$ suggests a transition between two types of instabilities.
The electrostatic potential along the parallel coordinate is shown in figure~\ref{fig:field} for two unstable modes at $k_y\rho_s=0.2$ and $k_y\rho_s=0.4$.  
Although the modes exhibit ballooning characteristics, i.e. the perturbation $\delta\phi$ remains relatively localized around $z = 0$, some oscillatory structures extend significantly along the $z$ direction. Consequently, resolving these modes accurately requires a large value of $n_x$, which, in turn, corresponds to an extended parallel domain in ballooning space. 

\begin{figure}
\centering
    \subfloat[]{\includegraphics[width=0.49\linewidth]{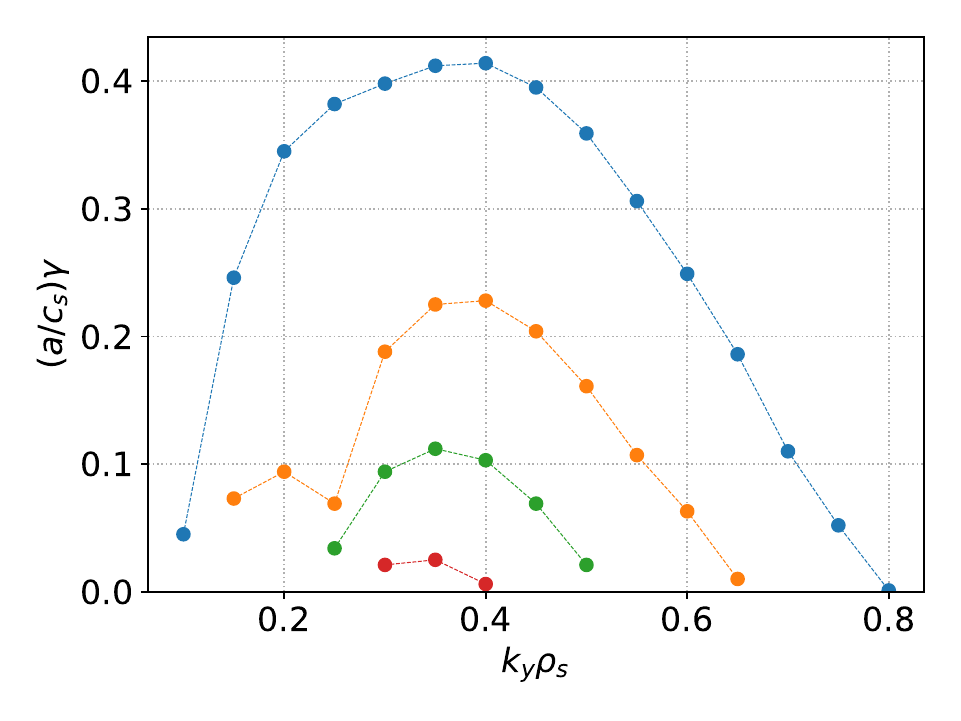}}
    \subfloat[]{\includegraphics[width=0.49\linewidth]{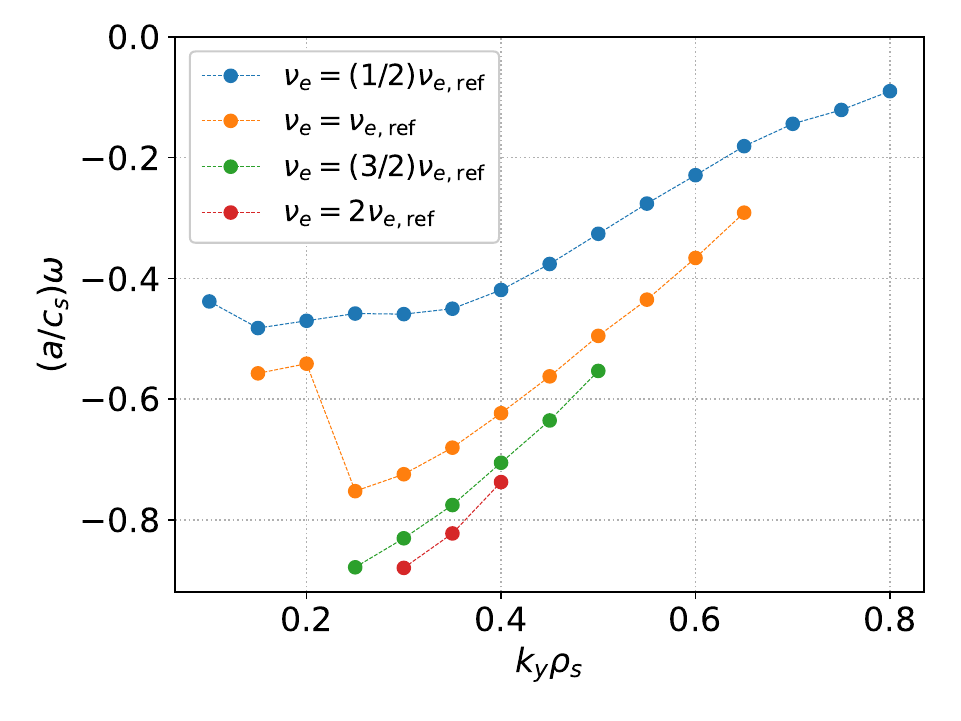}}
    \caption{Growth rate (a) and real frequency (b) as functions of $k_y\rho_s$ from GENE local linear simulations at three different values of electron collision frequency. The reference value of the electron collision frequency, which is directly evaluated from the electron temperature and density profiles of the GBS reference simulation, is $\nu_{e, \mathrm{ref}} \simeq 0.6\ c_s/a$. Only unstable modes are shown.}
    \label{fig:growth_rate}
\end{figure}

The instability is electrostatic, as the amplitude of \(\delta A_{\parallel}\) is negligible in comparison with that of \(\delta \phi\). Moreover, the extended oscillations along the parallel direction and the increasing negative mode frequency with $k_y$ show similarities with the so-called ubiquitous mode~\citep{coppi1974,coppi1977}, which is a branch of collisionless trapped electron modes (TEMs). This observation is supported by the electron collision frequency scan presented in figure~\ref{fig:growth_rate}(a), which shows a pronounced reduction in the growth rate across all $k_y$ modes as the electron collision frequency increases.
We emphasize that unstable ubiquitous modes, and more generally TEMs, have also been identified in the outer core region of MST RFP plasmas~\citep{carmody2015}. 

\begin{figure}
\centering
    \subfloat[]{\includegraphics[width=0.49\linewidth]{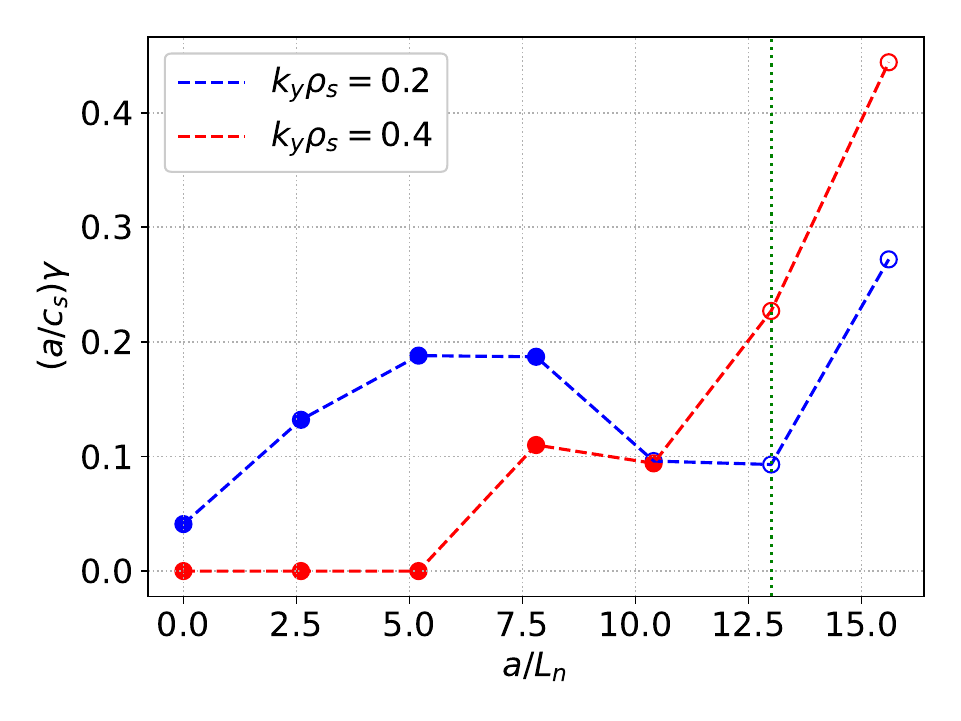}}
    \subfloat[]{\includegraphics[width=0.49\linewidth]{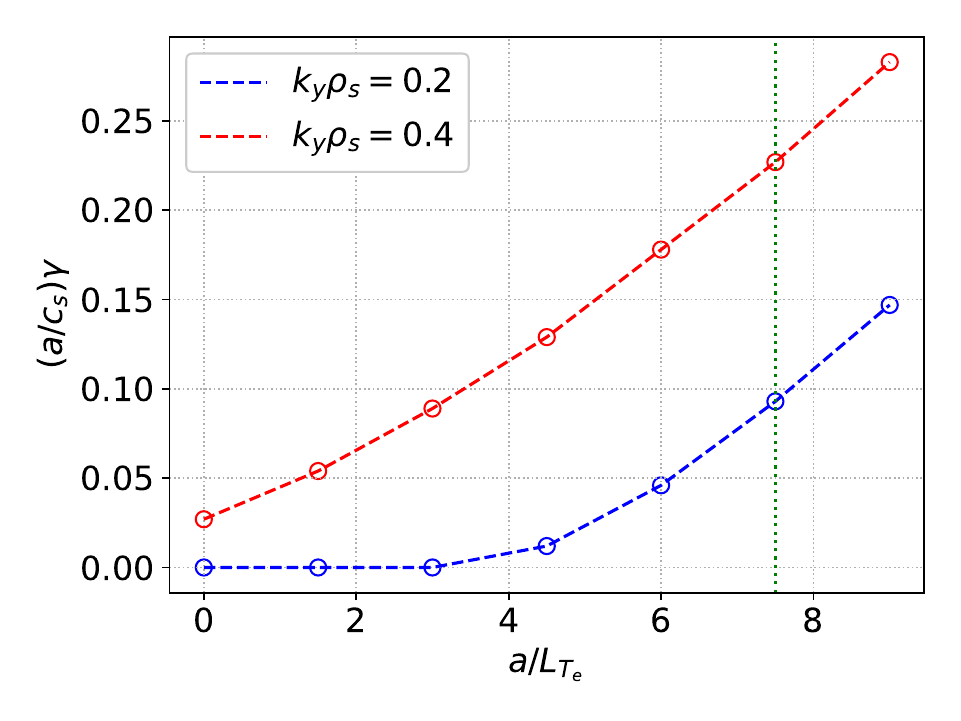}}
    \caption{Growth rate of the $k_y\rho_s=0.2$ (blue markers) and $k_y\rho_s=0.4$ (red markers) modes as a function of the normalized electron density (a) and temperature (b) gradients. The open (solid) markers indicate negative (positive) mode frequency. The growth rate of stable modes is set to zero. The vertical dotted lines indicate the reference value of the normalized electron density and temperature  gradients.}
    \label{fig:scans}
\end{figure}

Figure~\ref{fig:scans} shows the growth rate at $k_y\rho_s=0.2$ and $k_y\rho_s=0.4$ as a function of the normalized electron density and temperature gradients. The growth rate increases with both the density and temperature gradients. A transition to an unstable mode with positive frequency is observed at reduced density gradient at $k_y\rho_s=0.2$. This is an ion temperature gradient (ITG) mode.  
A similar transition occurs at $k_y\rho_s=0.4$, although this mode is stabilized below $a/L_n\simeq 6$. The presence of this hybrid ITG/TEM instability is in agreement with previous linear gyrokinetic studies carried out in RFP plasmas~\citep{carmody2013,liu2014,predebon2015,carmody2015}.
We also observe that the $k_y\rho_s = 0.4$ mode remains unstable over the entire $a/L_{T_e}$ range considered in figure~\ref{fig:scans}(b), including the case with vanishing electron temperature gradient. In contrast, the $k_y\rho_s = 0.2$ mode is stabilized for $a/L_{T_e} \lesssim 4$, highlighting a qualitative difference between these two modes, consistent with the distinct $k_y$ dependence of the mode frequency shown in figure~\ref{fig:growth_rate}(b).   

This linear gyrokinetic analysis points out the important role played by  trapped electrons in the edge of the considered RFP plasmas. Their dynamics is not incorporated in the two-fluid model implemented in GBS, which consequently limits its applicability for accurately describing turbulence in the edge region of RFP plasmas.
It is interesting to note that TEMs have also been recently found in gyrokinetic tokamak SOL turbulence simulations~\citep{frei2025}, thus providing further evidence of the similarity between tokamak and RFP turbulence in the plasma boundary region. 

\section{Conclusions}
\label{sec:conclusion}

This work presents the results of a set of flux-driven global turbulence simulations carried out with the GBS code in a RFP plasma, including the reversal surface and magnetic chaos generated by overlapping magnetic islands.
Turbulence simulations, carried out at various values of $\nu_0$ and $\beta_{e0}$, indicate that RBM-driven turbulence develops within a narrow radial layer localized around the reversal surface.
Near the vessel wall and in proximity of the reversal surface, the pressure profile is flattened by the presence of $m=0$ magnetic islands. In the outer core region, turbulence develops only at very large values of $\nu_0$ or $\beta_{e0}$.

The theoretical framework developed in \citet{giacomin2022turbulent} to characterize different edge turbulent transport regimes as functions of $\nu_0$ and $\beta_{e0}$ is extended herein to the RFP boundary plasma. This extension is achieved by modifying the theoretical scaling of the equilibrium pressure gradient length, originally derived in \citet{giacomin2020transp} for the tokamak edge,  to account for the different parallel connection length of RFP configurations.  
The modified scaling law of the pressure gradient length, Eq.~\eqref{eqn:lp_rfp}, leads to new boundaries of density and $\beta$ limits, shown in figure~\ref{fig:phase_space}.
Despite the lack of a favorable magnetic curvature region in RFP plasmas, the parallel correlation length of RBM-driven turbulent fluctuations remains shorter than the characteristic values observed in tokamak configurations (see figure~\ref{fig:lpar}(b)). As a result, the plasma collisionality, quantified by the ratio $L_\parallel/\lambda_e$, does not exceed values of order unity. The low collisionality constrains the applicability of two-fluid model to a narrow plasma region in the edge, where RBMs are driven. 
Conversely, the weak cross-field transport induced by RBMs represents a subdominant transport channel and reduces the strong confinement–resistivity correlation that is typically observed in tokamak boundary turbulence simulations, except when unrealistically large values of $\nu_0$ are considered. In particular, the pressure gradient scale length in the edge region is set by turbulent transport only for unphysically large values of $\nu_0$. 

The observed agreement between the numerically obtained and experimentally measured power spectra supports the presence of unstable RBMs in RFP configurations, but do not exclude the presence of kinetic instabilities that may contribute to cross-field turbulent transport. Despite the limitations of a fluid approach, GBS simulations reproduce the low-$k_\varphi$ resistive modes observed in the experiment and provide useful insights on the role of resistive turbulence in RFP plasmas. 
On the other hand, the present work suggests that a comprehensive description of turbulent transport in the boundary region of RFP plasmas may require full-$F$ global gyrokinetic turbulence simulations. 
This conclusion is also supported by local linear gyrokinetic simulations performed at $r/a = 0.8$ of the corresponding RFP axisymmetric magnetic equilibrium. In fact, despite the limits of the local approach, the simple magnetic geometry (circular) and the absence of magnetic chaos, these linear simulations reveal the presence of unstable trapped electron modes that lie beyond the descriptive capabilities of the fluid model implemented in GBS. 
Additional investigations will be conducted in future works to more rigorously assess the role of kinetic effects in determining RFP boundary turbulence.

\section*{Acknowledgments}
The authors would like to thank P. Ricci and the GBS team for useful discussions. This work has been carried out within the framework of Italian National Recovery and Resilience Plan (NRRP), funded by the European Union - NextGenerationEU (Mission 4, Component 2, Investment 3.1 - Area ESFRI Energy - Call for tender No. 3264 of 28-12-2021 of Italian University and Research Ministry (MUR), Project ID IR0000007, MUR Concession Decree No. 243 del 04/08/2022, CUP B53C22003070006, "NEFERTARI – New Equipment for Fusion Experimental Research and Technological Advancements with Rfx Infrastructure").
Views and opinions expressed are however those of the author(s) only and do not necessarily reflect those of the European Union or the European Commission. Neither the European Union nor the European Commission can be held responsible for them.
We acknowledge EuroHPC JU for awarding the project ID EHPC-REG-2024R02-031 access to Discoverer HPC (Sofia Tech Park, Bulgaria).

\appendix

\section{Differential operator coefficients}
\label{app:coeff}
Coefficients of the parallel Laplacian operator (see Eq.~\eqref{eqn:lapl_par}):
\begin{equation}
    \begin{split}
        a_R =\, &\frac{1}{B^4} \Bigl(B_Z  \bigl(\pdz B_R  (B^2 -B_R ^2)-B_R ^2 \pdr B_Z \bigr) +B_R  \pdr B_R  (B^2 -B_R ^2)\\&-B_R B_Z ^2 \pdz B_Z \Bigr) +\frac{B_\varphi}{B^4 R}  \Bigl(\pdp B_R  (B^2 -B_R ^2) \\&-B_R  \bigl(B_R  \pdr B_\varphi  R+B_Z  (\pdp B_Z +\pdz B_\varphi R)\bigr)\Bigr) -\frac{B_R  B_\varphi ^2}{B^4 R} \pdp B_\varphi 
    \end{split}
\end{equation}
\begin{equation}
    \begin{split}
        a_Z =\, &-\frac{B_R}{B^4 R}  \Bigl(-B^2  \pdr B_Z  R+B_Z ^2 R (\pdz B_R +\pdr B_Z ) +B_Z  B_\varphi  (\pdp B_R +\pdr B_\varphi  R)\Bigl)\\        
        &+\frac{B_\varphi}{B^4 R}  \Bigl(\pdp B_Z (B^2 -B_Z ^2)-B_Z ^2 \pdz B_\varphi  R\Bigr)+ \frac{1}{B^4 R} \Bigl(B_Z  \pdz B_Z  R (B^2 -B_Z ^2)\\
   &-B_R ^2 B_Z \pdr B_R  R-B_Z B_\varphi ^2 \pdp B_\varphi \Bigr)
    \end{split}
\end{equation}
\begin{equation}
    \begin{split}
        a_\varphi =\, & \frac{B_R}{B^4 R^2}  \Bigl(\pdr B_\varphi  R (B^2 -B_\varphi ^2)-B_\varphi  \bigl(B^2 +B_Z  R (\pdz B_R +\pdr B_Z )+B_\varphi  \pdp B_R \bigr)\Bigr)\\
        &-\frac{B_\varphi}{B^4 R^2} \Bigl(\pdp B_\varphi  (B_\varphi ^2-B^2 )+B_Z ^2 \pdz B_Z  R+B_Z  B_\varphi  \pdp B_Z \Bigr)\\&+ \frac{B_Z}{B^4 R^2}  \pdz B_\varphi  R (B^2 -B_\varphi ^2)-\frac{B_R^2 B_\varphi}{B^4 R^2} \pdr B_R  R
    \end{split}
\end{equation}
\begin{align}
    a_{RR} =& \frac{B_R^2}{B^2}\\
    a_{ZZ} =& \frac{B_Z^2}{B^2}\\
    a_{\varphi\varphi} =& \frac{B_\varphi^2}{B^2}
\end{align}
\begin{align}
    a_{RZ} =& \frac{2B_RB_Z}{B^2}\\
    a_{R\varphi} =& \frac{2B_RB_\varphi}{B^2}\\
    a_{Z\varphi} =& \frac{2B_ZB_\varphi}{B^2}
\end{align}

Coefficients of the perpendicular Laplacian operator (see Eq.~\eqref{eqn:lapl_perp}):
\begin{equation}
\begin{split}
    c_R = \frac{1}{{B^4 R}}\Bigl(&B_\varphi  \bigl(2R B^2  \pdr B_\varphi  -\pdp B_R  (B^2 -2 B_R ^2)\bigr) +B_Z ^2 \bigl(RB^2 \\ &+2 B_R (\pdz B_Z -\pdr B_R )-2 B_\varphi  \pdr B_\varphi  R\bigr)+B_Z  \bigl(2R \pdr B_Z   (B^2 -B_\varphi ^2)\\
   &-RB^2\pdz B_R  +2 B_R  (RB_R  \pdz B_R  +B_\varphi  (\pdp B_Z +R\pdz B_\varphi  ))\bigr)\\
   &+B_\varphi ^2 \bigl(B^2 +2 B_R 
   (\pdp B_\varphi -R\pdr B_R  )\bigr) -B^2  B_R  (R\pdz B_Z  +\pdp B_\varphi )\\&-2R B_Z ^3 \pdr B_Z  -2 RB_\varphi ^3 \pdr B_\varphi\Bigl)
   \end{split}
\end{equation}
\begin{equation}
\begin{split}
    c_Z = \frac{1}{B^4 R}\Bigl(&B_R  \bigl(B_Z  (2 B_\varphi  (\pdp B_R +R\pdr B_\varphi  )-B^2 ) \\&+2R \pdz B_R (B^2 -B_\varphi ^2)-RB^2\pdr B_Z +2R B_Z ^2 \pdr B_Z  \bigr) \\&+ B_\varphi  \bigl(2R B^2  \pdz B_\varphi -\pdp B_Z  (B^2 -2 B_Z ^2)\bigr)-B^2  B_Z  (R\pdr B_R  +\pdp B_\varphi )\\&-2R B_R ^3 \pdz B_R -2R B_R ^2 \bigl(B_Z  (\pdz B_Z -\pdr B_R )+B_\varphi  \pdz B_\varphi \bigr)\\
    &+2 B_Z  B_\varphi ^2 (\pdp B_\varphi -R\pdz B_Z )-2R B_\varphi ^3 \pdz B_\varphi  \Bigr)
   \end{split}
\end{equation}
\begin{equation}
\begin{split}
    c_\varphi = \frac{1}{B^4 R^2}\Bigr(&B_R  (2 \pdp B_R  (B^2 -B_Z ^2)-R\pdr B_\varphi  (B^2 -2 B_\varphi ^2)\\&+2 RB_Z  B_\varphi (\pdz B_R +\pdr B_Z ))+B_Z  (2 B^2  \pdp B_Z \\&-R\pdz B_\varphi   (B^2 -2 B_\varphi ^2))-RB^2  B_\varphi   (\pdr B_R +\pdz B_Z )\\-&2 B_R ^3 \pdp B_R -2 B_R ^2 (B_Z \pdp B_Z +B_\varphi  (\pdp B_\varphi -R\pdr B_R ))\\-&2 B_Z ^3 \pdp B_Z -2 B_Z ^2 B_\varphi  (\pdp B_\varphi -R\pdz B_Z )\Bigl)
\end{split}
\end{equation}
\begin{align}
    c_{RR} &= \frac{B_Z^2+B_\varphi^2}{B^2}\\
    c_{ZZ} &= \frac{B_R^2+B_\varphi^2}{B^2}\\
    c_{\varphi\varphi} &= \frac{B_R^2+B_Z^2}{B^2}\\
    c_{R Z} &= -\frac{2B_RB_Z}{B^2R}\\
    c_{R\varphi} &= -\frac{2B_RB_\varphi}{B^2R}\\
    c_{Z\varphi} &= -\frac{2B_ZB_\varphi}{B^2R}
\end{align}

The modified Laplacian operator in the right-hand side of Poisson law, Eq.~\eqref{eqn:poisson}, is written as
\begin{equation}
\begin{split}
    \nabla \cdot(n\nabla_\perp \phi) =\,& d_R\pdr \phi +d_Z \pdz \phi + d_\varphi \partial_\varphi\phi + d_{RR} \pdr^2\phi\\&+d_{ZZ}\pdz^2\phi+ d_{\varphi\varphi}\pdp^2\phi+d_{RZ}\partial_{RZ}\phi\\&+d_{R\varphi}\partial_{R\varphi}\phi+d_{Z\varphi}\partial_{R\varphi}\phi\,,
\end{split}
\end{equation}
with
\begin{equation}
\begin{split}
    d_R =\, & \frac{B_Z}{B^6 R}  \Bigl(B_R  \bigl(4 B_\varphi  n (\pdp B_Z +\pdz B_\varphi  R)-B^2  \pdz n  R\bigr)\\&+n R \bigl(2\pdr B_Z  (B^2 -2 B_\varphi ^2)-B^2  \pdz B_R \bigr)+4 B_R ^2 \pdz B_R  n R\Bigr) \\
    &-\frac{B_\varphi}{B^6 R}  \Bigl(\pdp B_R  n (B^2 -4 B_R ^2) +B^2  (B_R  \pdp n -2
   \pdr B_\varphi  n R)\Bigl)\\
   &+\frac{B_Z^2}{B^6 R} \Bigl(n (B^2 +4 B_R  R (\pdz B_Z -\pdr B_R ) -4 B_\varphi \pdr B_\varphi  R) +B^2  \pdr n  R\Bigr) \\
   &+\frac{B_\varphi^2}{B^6 R} \Bigl(B^2  (R\pdr n +n) +4 B_R  n (\pdp B_\varphi -\pdr B_R  R)\Bigr)\\
   &-\frac{1}{B^6 R}(B^2  B_R  n (\pdz B_Z  R+\pdp B_\varphi )+4 B_Z ^3 \pdr B_Z  n R + 4 B_\varphi ^3 \pdr B_\varphi  n R)
\end{split}
\end{equation}

\begin{equation}
\begin{split}
    d_Z =& -\frac{B_R^2}{B^6} \Bigl(-B^2  \pdz n +4 B_Z  n (\pdz B_Z -\pdr B_R )+4 B_\varphi  \pdz B_\varphi n\Bigr) \\
    &+\frac{B_R}{B^6 R}  \Bigl(B_Z  \bigl(n (B^2 -4 B_\varphi  (\pdp B_R +\pdr B_\varphi  R))+B^2  \pdr n  R\bigr)\\
    &+n R \bigl(B^2  \pdr B_Z -2 \pdz B_R  (B^2 -2 B_\varphi ^2)\bigr) -4 B_Z ^2 \pdr B_Z n R\Bigr) \\
    &+\frac{B_\varphi}{B^6 R}  \Bigl(\pdp B_Z  n (B^2 -4 B_Z ^2) +B^2  (B_Z  \pdp n -2
   \pdz B_\varphi  n R)\Bigr)\\   
   &-\frac{B_\varphi^2}{B^6 R} \Bigl(B^2  \pdz n  R+4 B_Z  n (\pdp B_\varphi -\pdz B_Z R)\Bigr)\\
   &+\frac{1}{B^6 R}\Bigl(B^2  B_Z  n (\pdr B_R  R+\pdp B_\varphi )+4 B_R ^3 \pdz B_R  n R+4 B_\varphi ^3 \pdz B_\varphi n R\Bigr)
\end{split}
\end{equation}

\begin{equation}
\begin{split}
    d_\varphi =\,& \frac{B_R^2}{B^6 R^2} \Bigl(B^2  \pdp n -4 n \bigl(B_Z  \pdp B_Z +B_\varphi  (\pdp B_\varphi -\pdr B_R R)\bigr)\Bigr)\\
    &+\frac{B_R}{B^6 R^2}  \Bigl(2 \pdp B_R  n (B^2 -2 B_Z ^2)+R \bigl(-\pdr B_\varphi  n (B^2 -4 B_\varphi ^2)\\
    &-B^2  B_\varphi  \pdr n +4 B_Z  B_\varphi  n
   (\pdz B_R +\pdr B_Z )\bigr)\Bigr)\\
   &+\frac{B_Z^2}{B^6 R^2} \Bigl(B^2  \pdp n -4 B_\varphi  n
   (\pdp B_\varphi -\pdz B_Z  R)\Bigr)\\
   &+\frac{B_Z}{B^6 R^2} \Bigl(2 B^2  \pdp B_Z  n-R (\pdz B_\varphi  n (B^2 -4
   B_\varphi ^2) +B^2  B_\varphi  \pdz n )\Bigr) \\
   &-\frac{1}{B^6 R^2} \Bigl(B^2  B_\varphi  n R (\pdr B_R +\pdz B_Z )+4 B_R ^3 \pdp B_R  n+4 B_Z ^3 \pdp B_Z  n\Bigr)
\end{split}
\end{equation}

\begin{align}
    d_{RR} &= \frac{n}{B^2}\Bigl(B_Z^2 + B_\varphi^2\Bigr)\\
    d_{ZZ} &= \frac{n}{B^2}\Bigl(B_R^2 + B_\varphi^2\Bigr)\\
    d_{\varphi\varphi} &= \frac{n}{R^2B^2}\Bigl(B_R^2 + B_Z^2\Bigr)\\
    d_{RZ} &= -\frac{2nB_R B_Z}{RB^2}\\
    d_{R\varphi} &= -\frac{2nB_R B_\varphi}{RB^2}\\
    d_{Z\varphi} &= -\frac{2nB_Z B_\varphi}{RB^2}
\end{align}

\section{Resolution convergence study of linear gyrokinetic simulations}
\label{app:resolution}

\begin{figure}
\centering
    \subfloat[]{\includegraphics[width=0.49\linewidth]{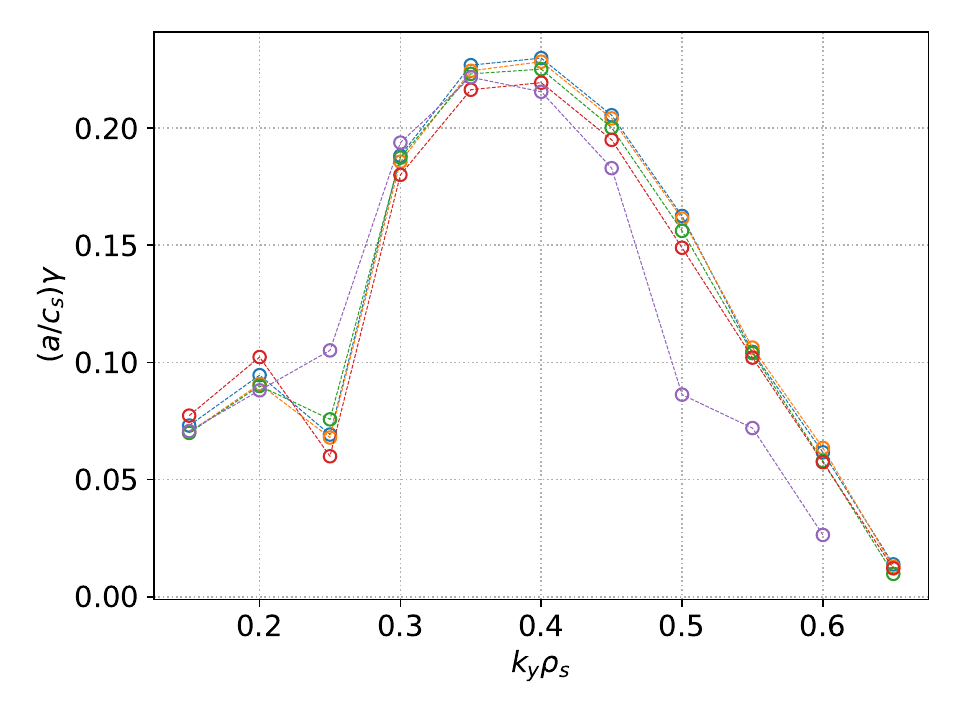}}
    \subfloat[]{\includegraphics[width=0.49\linewidth]{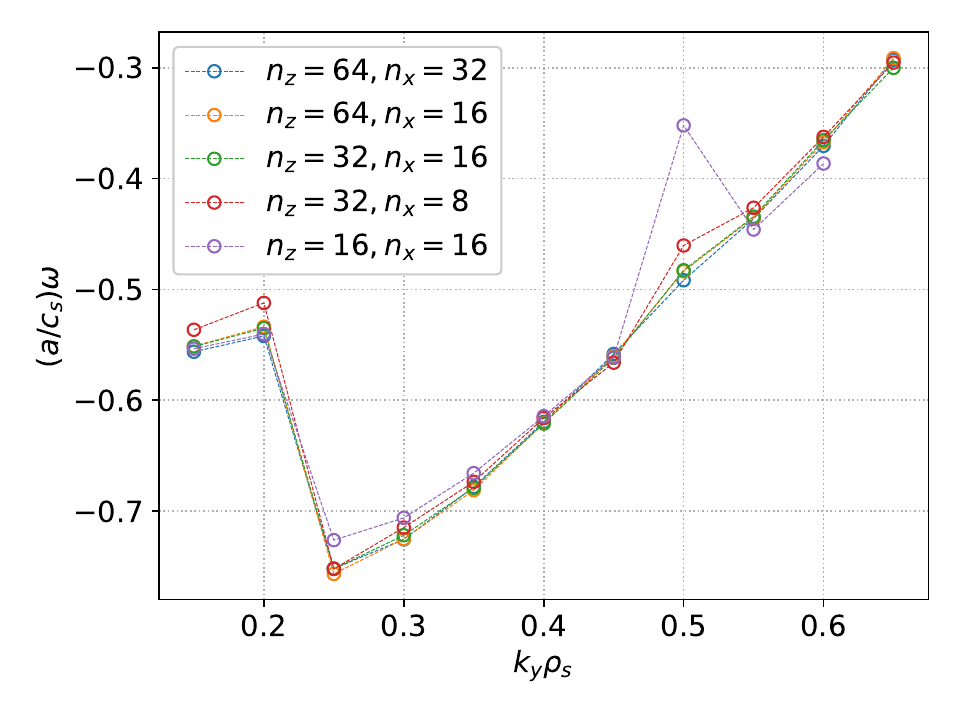}}
    \caption{Growth rate (a) and mode frequency (b) as functions of $k_y\rho_s$ from a set of GENE local linear simulations with various radial and parallel grid resolutions.}
    \label{fig:resolution}
\end{figure}

A set of GENE local linear simulations are carried out at increasing radial and parallel grid resolutions to check the numerical convergence of the linear gyrokinetic results reported in section~\ref{sec:gyrokinetic}. The growth rate and mode frequency $k_y$ spectra from this set of simulations are shown in figure~\ref{fig:resolution}.
Numerical convergence of growth rate and mode frequency is achieved for $n_z\geq 64$ and $n_x\geq 16$.

Although the growth rate and mode frequency obtained in the simulation with $n_z=64$ and $n_x=16$ are nearly identical to those from the simulation with $n_z=64$ and $n_x=32$, accurate resolution of the extended oscillatory structure of $\delta \phi$ along the parallel direction (see figure~\ref{fig:field}) requires a sufficiently extended ballooning-space domain. This, in turn, requires a value of $n_x$ larger than the minimum imposed solely by the numerical convergence of $\gamma$ and $\omega$. For this reason, the simulations in section~\ref{sec:gyrokinetic} are carried out with $n_x=32$.

\bibliographystyle{jpp}
\bibliography{bibliography}

@article{Ricci2012,
author = {Ricci, P. and Halpern, F. D. and Jolliet, S. and Loizu, J. and Mosetto, A. and Fasoli, A. and Furno, I. and Theiler, C.},
doi = {10.1088/0741-3335/54/12/124047},
issn = {0741-3335},
journal = {Plasma Physics and Controlled Fusion},
number = {12},
pages = {124047},
title = {{Simulation of plasma turbulence in scrape-off layer conditions: the GBS code, simulation results and code validation}},
volume = {54},
year = {2012}
}

@article{Zeiler1997,
author = {Zeiler, A. and Drake, J. F. and Rogers, B.},
doi = {10.1063/1.872368},
issn = {1070664X},
journal = {Physics of Plasmas},
number = {6},
pages = {2134--2138},
title = {{Nonlinear reduced Braginskii equations with ion thermal dynamics in toroidal plasma}},
volume = {4},
year = {1997}
}

@article{Mosetto2013,
author = {Mosetto, A. and Halpern, F. D. and Jolliet, S. and Loizu, J. and Ricci, P.},
doi = {10.1063/1.4821597},
isbn = {9781632663108},
issn = {1070-664X},
journal = {Physics of Plasmas},
number = {9},
pages = {092308},
title = {{Turbulent regimes in the tokamak scrape-off layer}},
url = {http://aip.scitation.org/doi/10.1063/1.4821597},
volume = {20},
year = {2013}
}

@article{giacomin2020snow,
  title={{Turbulence and flows in the plasma boundary of snowflake magnetic configurations}},
  author={Giacomin, M. and Stenger, L.N. and Ricci, P.},
  journal={Nuclear Fusion},
  volume={60},
  number={2},
  pages={024001},
  year={2020},
  publisher={IOP Publishing}
}

@article{riva2014,
  title={Verification methodology for plasma simulations and application to a scrape-off layer turbulence code},
  author={Riva, Fabio and Ricci, Paolo and Halpern, Federico D. and Jolliet, S{\'e}bastien and Loizu, Joaquim and Mosetto, Annamaria},
  journal={Physics of Plasmas},
  volume={21},
  number={6},
  pages={062301},
  year={2014},
  publisher={American Institute of Physics}
}

@article{giacomin2020transp, 
title={Investigation of turbulent transport regimes in the tokamak edge by using two-fluid simulations}, 
volume={86}, 
DOI={10.1017/S0022377820000914}, 
number={5}, 
journal={Journal of Plasma Physics}, 
publisher={Cambridge University Press}, 
author={Giacomin, M. and Ricci, P.}, 
year={2020}, 
pages={905860502}
}

@article{giacomin2021,
  title={Theory-based scaling laws of near and far scrape-off layer widths in single-null L-mode discharges},
  author={Giacomin, M. and Stagni, A. and Ricci, P. and Boedo, J. A. and Horacek, J. and Reimerdes, H. and Tsui, C. K.},
  journal={Nuclear Fusion},
  year={2021},
  volume={61},
  pages={076002},
  number={7},
  publisher={IOP Publishing}
}

@article{giacomin2022gbs,
  title={{The GBS code for the self-consistent simulation of plasma turbulence and kinetic neutral dynamics in the tokamak boundary}},
  author={Giacomin, M. and Ricci, P. and Coroado, A. and Fourestey, G. and Galassi, D. and Lanti, E. and Mancini, D. and Richart, N. and Stenger, L. N. and Varini, N.},
  journal={Journal of Computational Physics},
  volume={463},
  pages={111294},
  year={2022},
  publisher={Elsevier}
}

@article{giacomin2022density,
  title = {{First-Principles Density Limit Scaling in Tokamaks Based on Edge Turbulent Transport and Implications for ITER}},
  author = {Giacomin, M. and Pau, A. and Ricci, P. and Sauter, O. and Eich, T. and the ASDEX Upgrade team and JET Contributors and the TCV team},
  journal = {Phys. Rev. Lett.},
  volume = {128},
  issue = {18},
  pages = {185003},
  year = {2022},
  publisher = {American Physical Society},
  doi = {10.1103/PhysRevLett.128.185003},
}

@article{giacomin2022turbulent,
  title={Turbulent transport regimes in the tokamak boundary and operational limits},
  author={Giacomin, M. and Ricci, P.},
  journal={Physics of Plasmas},
  volume={29},
  number={6},
  year={2022},
  publisher={AIP Publishing}
}

@article{lim2023,
  title={{Effect of triangularity on plasma turbulence and the SOL-width scaling in L-mode diverted tokamak configurations}},
  author={Lim, K. and Giacomin, M. and Ricci, P. and Coelho, A. and F{\'e}vrier, O. and Mancini, D. and Silvagni, D. and Stenger, L.},
  journal={Plasma Physics and Controlled Fusion},
  volume={65},
  number={8},
  pages={085006},
  year={2023},
  publisher={IOP Publishing}
}

@article{coelho2022,
  title={Global fluid simulation of plasma turbulence in a stellarator with an island divertor},
  author={Coelho, A. J. and Loizu, J. and Ricci, P. and Giacomin, M.},
  journal={Nuclear Fusion},
  volume={62},
  number={7},
  pages={074004},
  year={2022},
  publisher={IOP Publishing}
}

@article{oliveira2022,
  title={{Validation of edge turbulence codes against the TCV-X21 diverted L-mode reference case}},
  author={Oliveira, D. S. and Body, T. and Galassi, D. and Theiler, C. and Laribi, E. and Tamain, P. and Stegmeir, A. and Giacomin, M. and Zholobenko, W. and Ricci, P. and others},
  journal={Nuclear Fusion},
  volume={62},
  number={9},
  pages={096001},
  year={2022},
  publisher={IOP Publishing}
}

@article{spolaore2017,
  title={{H-mode achievement and edge features in RFX-mod tokamak operation}},
  author={Spolaore, M. and Cavazzana, R. and Marrelli, L. and Carraro, L. and Franz, P. and Spagnolo, S. and Zaniol, B. and Zuin, M. and Cordaro, L. and Dal Bello, S. and others},
  journal={Nuclear Fusion},
  volume={57},
  number={11},
  pages={116039},
  year={2017},
  publisher={IOP Publishing}
}

@article{braginskii1965,
  title={Transport processes in a plasma},
  author={Braginskii, S. I.},
  journal={Reviews of Plasma Physics},
  volume={1},
  pages={205},
  year={1965}
}

@article{marrelli2021,
  title={The reversed field pinch},
  author={Marrelli, L. and Martin, P. and Puiatti, M. E. and Sarff, J. S. and Chapman, B. E. and Drake, J. R. and Escande, D. F. and Masamune, S.},
  journal={Nuclear Fusion},
  volume={61},
  number={2},
  pages={023001},
  year={2021},
  publisher={IOP Publishing}
}

@article{lorenzini2009,
  title={Self-organized helical equilibria as a new paradigm for ohmically heated fusion plasmas},
  author={Lorenzini, Rita and Martines, E and Piovesan, P and Terranova, D and Zanca, P and Zuin, M and Alfier, A and Bonfiglio, D and Bonomo, F and Canton, A and others},
  journal={Nature Physics},
  volume={5},
  number={8},
  pages={570--574},
  year={2009},
  publisher={Nature Publishing Group UK London}
}

@article{zanca2004,
  title={Reconstruction of the magnetic perturbation in a toroidal reversed field pinch},
  author={Zanca, P. and Terranova, D.},
  journal={Plasma Physics and Controlled Fusion},
  volume={46},
  number={7},
  pages={1115},
  year={2004},
  publisher={IOP Publishing}
}

@article{zuin2013,
  title={Experimental observation of microtearing modes in a toroidal fusion plasma},
  author={Zuin, M. and Spagnolo, S. and Predebon, I. and Sattin, F and Auriemma, F. and Cavazzana, R. and Fassina, A. and Martines, E. and Paccagnella, R. and Spolaore, M. and others},
  journal={Physical Review Letters},
  volume={110},
  number={5},
  pages={055002},
  year={2013},
  publisher={APS}
}

@article{giacomin2025,
  title={{Three-dimensional boundary turbulence simulations of a RFX-mod plasma in the presence of voltage biasing}},
  author={Giacomin, M. and Vianello, N. and Cavazzana, R. and Molisani, S. and Spolaore, M. and Zuin, M.},
  journal={Nuclear Fusion},
  volume={65},
  number={3},
  pages={036013},
  year={2025},
  publisher={IOP Publishing}
}

@article{vianello2016,
  title={{On the statistics and features of turbulent structures in RFX-mod}},
  author={Vianello, N. and Spolaore, M. and Agostini, M. and Cavazzana, R. and De Masi, G. and Martines, E. and Momo, B. and Scarin, P. and Spagnolo, S. and Zuin, M.},
  journal={Plasma Physics and Controlled Fusion},
  volume={58},
  number={4},
  pages={044009},
  year={2016},
  publisher={IOP Publishing}
}

@article{predebon2010,
  title={Microtearing modes in reversed field pinch plasmas},
  author={Predebon, I. and Sattin, F. and Veranda, M. and Bonfiglio, D. and Cappello, S.},
  journal={Physical Review Letters},
  volume={105},
  number={19},
  pages={195001},
  year={2010},
  publisher={APS}
}

@article{rea2015,
  title={{Comparative studies of electrostatic turbulence induced transport in presence of resonant magnetic perturbations in RFX-mod}},
  author={Rea, C. and Vianello, N. and Agostini, M. and Cavazzana, R. and De Masi, G. and Martines, E. and Momo, B. and Scarin, P. and Spagnolo, S. and Spizzo, G. and others},
  journal={Nuclear Fusion},
  volume={55},
  number={11},
  pages={113021},
  year={2015},
  publisher={IOP Publishing}
}

@article{carmody2013,
  title={Gyrokinetic studies of microinstabilities in the reversed field pinch},
  author={Carmody, D. and Pueschel, M. J. and Terry, P. W.},
  journal={Physics of Plasmas},
  volume={20},
  number={5},
  year={2013},
  publisher={AIP Publishing}
}

@article{predebon2015,
  title={{Ion temperature gradient turbulence in helical and axisymmetric RFP plasmas}},
  author={Predebon, I and Xanthopoulos, P},
  journal={Physics of Plasmas},
  volume={22},
  number={5},
  year={2015},
  publisher={AIP Publishing}
}

@article{williams2017,
  title={{Turbulence, transport, and zonal flows in the Madison symmetric torus reversed-field pinch}},
  author={Williams, Z. R. and Pueschel, M. J. and Terry, P. W. and Hauff, T.},
  journal={Physics of Plasmas},
  volume={24},
  number={12},
  year={2017},
  publisher={AIP Publishing}
}

@article{thuecks2017,
  title={Evidence for drift waves in the turbulence of reversed field pinch plasmas},
  author={Thuecks, D. J. and Almagri, A. F. and Sarff, J. S. and Terry, P. W.},
  journal={Physics of Plasmas},
  volume={24},
  number={2},
  year={2017},
  publisher={AIP Publishing}
}

@article{ren2011,
  title={Experimental observation of anisotropic magnetic turbulence in a reversed field pinch plasma},
  author={Ren, Y. and Almagri, A. F. and Fiksel, G. and Prager, S. C. and Sarff, J. S. and Terry, P. W.},
  journal={Physical Review Letters},
  volume={107},
  number={19},
  pages={195002},
  year={2011},
  publisher={APS}
}

@article{mancini2023,
  title={{Self-consistent multi-component simulation of plasma turbulence and neutrals in detached conditions}},
  author={Mancini, D. and Ricci, P. and Vianello, N. and Van Parys, G. and Oliveira, D. S.},
  journal={Nuclear Fusion},
  volume={64},
  number={1},
  pages={016012},
  year={2023},
  publisher={IOP Publishing}
}

@Misc{petsc-web-page,
author = {S. Balay and S. Abhyankar and M. F. Adams and J. Brown and P. Brune and K. Buschelman and L. Dalcin and A. Dener and V. Eijkhout and W.~D. Grop and D. Karpeyev and D. Kaushik and M.~G. Knepley and D.~A. May and L. C. McInnes and others},
title =  {{PETS}c {W}eb page},
url =    {https://www.mcs.anl.gov/petsc},
howpublished = {\url{https://www.mcs.anl.gov/petsc}},
year = {2019}
}

@inproceedings{hypre,
  title={{hypre: A library of high performance preconditioners}},
  author={Falgout, R. D. and Yang, U. M.},
  booktitle={International Conference on computational science},
  pages={632--641},
  year={2002},
  organization={Springer}
}

@article{saad1993,
  title={{A flexible inner-outer preconditioned GMRES algorithm}},
  author={Saad, Y.},
  journal={SIAM Journal on Scientific Computing},
  volume={14},
  number={2},
  pages={461--469},
  year={1993},
  publisher={SIAM}
}

@article{peterson2013,
  title={Positivity preservation and advection algorithms with applications to edge plasma turbulence},
  author={Peterson, J. L. and Hammett, G. W.},
  journal={SIAM Journal on Scientific Computing},
  volume={35},
  number={3},
  pages={B576--B605},
  year={2013},
  publisher={SIAM}
}

@article{pitzal2023,
  title={{Landau-fluid simulations of edge-SOL turbulence with GRILLIX}},
  author={Pitzal, C. and Stegmeir, A. and Zholobenko, W. and Zhang, K. and Jenko, F.},
  journal={Physics of Plasmas},
  volume={30},
  number={12},
  year={2023},
  publisher={AIP Publishing}
}

@article{agostini2014,
  title={{Parallel and perpendicular structure of the edge turbulence in a three-dimensional magnetic field}},
  author={Agostini, M. and Scarin, P. and Spizzo, G. and Vianello, N. and Carraro, L.},
  journal={Plasma Physics and Controlled Fusion},
  volume={56},
  number={9},
  pages={095016},
  year={2014},
  publisher={IOP Publishing}
}

@article{giacomin2026,
  title={{Global flux-driven turbulence simulations in chaotic plasmas}},
  author={Giacomin, M. and Momo, B. and Predebon, I. and Vianello, N. and Zuin, M.},
  journal={Communications Physics},
  volume={9},
  number={249},
  pages={},
  year={2026},
  publisher={Nature Publishing Group UK London}
}

@article{zholobenko2021,
  title={{Electric field and turbulence in global Braginskii simulations across the ASDEX Upgrade edge and scrape-off layer}},
  author={Zholobenko, W. and Body, T. and Manz, P. and Stegmeir, A. and Zhu, B. and Griener, M. and Conway, G.D. and Coster, D. and Jenko, F. and ASDEX Upgrade Team},
  journal={Plasma Physics and Controlled Fusion},
  volume={63},
  number={3},
  pages={034001},
  year={2021},
  publisher={IOP Publishing}
}

@article{frei2025,
  title={{Turbulence and transport in spectrally accelerated full-f gyrokinetic simulations}},
  author={Frei, B.J. and Ulbl, P. and Pitzal, C. and Zholobenko, W. and Jenko, F.},
  journal={Nuclear Fusion},
  volume={65},
  number={11},
  pages={116026},
  year={2025},
  publisher={IOP Publishing}
}

@article{jenko2000,
  title={Electron temperature gradient driven turbulence},
  author={Jenko, F. and Dorland, W. and Kotschenreuther, M. and Rogers, B.N.},
  journal={Physics of Plasmas},
  volume={7},
  number={5},
  pages={1904--1910},
  year={2000},
  publisher={American Institute of Physics}
}

@article{miller1998,
  title={Noncircular, finite aspect ratio, local equilibrium model},
  author={Miller, R.L. and Chu, M.-S. and Greene, J. M. and Lin-Liu, Y.R. and Waltz, R.E.},
  journal={Physics of Plasmas},
  volume={5},
  number={4},
  year={1998}
}

@article{coppi1974,
  title={New trapped-electron instability},
  author={Coppi, Bruno and Rewoldt, Gregory},
  journal={Physical Review Letters},
  volume={33},
  number={22},
  pages={1329},
  year={1974},
  publisher={APS}
}

@article{carmody2015,
  title={Microturbulence studies of pulsed poloidal current drive discharges in the reversed field pinch},
  author={Carmody, D. and Pueschel, M.J. and Anderson, J.K. and Terry, P.W.},
  journal={Physics of Plasmas},
  volume={22},
  number={1},
  year={2015},
  publisher={AIP Publishing}
}

@article{jitsuk2026,
  title={{Turbulent Multiscale Interactions between Tearing Modes, Trapped-Electron Modes, and Zonal Flows}},
  author={Jitsuk, T. and Pueschel, M. J. and Terry, P. W. and Di Siena, A.},
  journal={Physical Review Letters},
  volume={136},
  number={1},
  pages={015101},
  year={2026},
  publisher={APS}
}

@article{predebon2019,
  title={{Electron temperature gradient driven instabilities in helical reversed-field pinch plasmas}},
  author={Predebon, I. and Xanthopoulos, P. and Gobbin, M.},
  journal={Plasma Physics and Controlled Fusion},
  volume={61},
  number={5},
  pages={055011},
  year={2019},
  publisher={IOP Publishing}
}

@article{duff2018,
  title={Observation of trapped-electron-mode microturbulence in reversed field pinch plasmas},
  author={Duff, J. R. and Williams, Z. R. and Brower, D. L. and Chapman, B. E. and Ding, W. X. and Pueschel, M. J. and Sarff, J. S. and Terry, P. W.},
  journal={Physics of Plasmas},
  volume={25},
  number={1},
  year={2018},
  publisher={AIP Publishing}
}

@article{coppi1977,
  title={Theory of the ubiquitous mode},
  author={Coppi, B. and Pegoraro, F.},
  journal={Nuclear Fusion},
  volume={17},
  number={5},
  pages={969--993},
  year={1977}
}

@article{liu2014,
  title={{Trapped electron effects on $\eta$ i-mode and trapped electron mode in RFP plasmas}},
  author={Liu, S. F. and Guo, S. C. and Kong, W. and Dong, J. Q.},
  journal={Nuclear Fusion},
  volume={54},
  number={4},
  pages={043006},
  year={2014},
  publisher={IOP Publishing}
}

\end{document}